\documentclass[1p, 12pt]{elsarticle}
\usepackage{amsmath}
\usepackage{amsfonts}
\usepackage{amssymb} 
\usepackage{amsthm}
\usepackage{mathtools}
\usepackage{mathabx}
\usepackage{xcolor}
\usepackage{colortbl}
\usepackage{graphicx}
\usepackage{float}
\usepackage{accents}
\usepackage{subcaption}
\usepackage{tabu}
\usepackage{tikz}
\usepackage{tkz-graph}
\usepackage{booktabs}
\usetikzlibrary{arrows, shapes, matrix, positioning, decorations.pathmorphing}

\newif\ifcomment
\commenttrue

\newtheorem{theorem}{Theorem}
\newtheorem{lemma}{Lemma}
\newtheorem{corollary}{Corollary}
\newtheorem{proposition}{Proposition}
\newtheorem{definition}{Definition}

\newcommand{\tuple}[1]{\langle #1 \rangle}

\newcommand{\Bset}[0]{\mathbb{B}}

\newcommand{\Nset}[0]{\mathbb{N}}

\newcommand{\partset}[1]{\ensuremath{\mathbf{2}^{#1}}}
\DeclareMathOperator{\dom}{\textbf{dom}}
\DeclareMathOperator{\codom}{\textbf{codom}}

\newcommand{\eqdef}{\ensuremath{\mathrel{\stackrel{\mathrm{def}}{=}}}}

\newcommand{\abbrev}[1]{#1, \relax}
\newcommand{\ie}[0]{\abbrev{\textit{i.e.}}}
\newcommand{\eg}[0]{\abbrev{\textit{e.g.}}}

\newcommand{\evo}[0]{\longrightarrow}
\newcommand{\xevo}[1]{\stackrel{#1}{\evo}}

\newcommand{\arc}{\ensuremath{\, \tikz[baseline=-0.5ex, scale=0.65, thin]{\draw[->, >=open triangle 45](0, 0)to[bend left=0] (1, 0);}\, }}
\newcommand{\blackarc}{\ensuremath{\, \tikz[baseline=-0.5ex, scale=0.65, thin]{\draw[->, >= triangle 45](0, 0)to[bend left=0] (1, 0);}\, }}
\newcommand{\sarc}[1]{\stackrel{\kern-1ex{#1}}{\arc}}
\newcommand{\sblackarc}[1]{\stackrel{\kern-1ex{#1}}{\blackarc}}

\newcommand{\shortblackarc}{\ensuremath{\, \tikz[baseline=-0.5ex, scale=0.45, thin]{\draw[->, >= triangle 45](0, 0)to[bend left=0] (1, 0);}\, }}

\newcommand{\sprt}[1]{\widehat{#1}}

\DeclareMathOperator{\minterm}{minterm}

\begin{document}
\makeatletter
\def\tabu@verticalmeasure{\everypar{}%
	\unless\ifnum\currentgrouptype=14 \let\tabu@currentgrouptype\currentgrouptype\fi
	\ifnum \tabu@currentgrouptype>12 
	\setbox\tabu@box =\hbox\bgroup
	\let\tabu@verticalspacing \tabu@verticalsp@lcr
	\d@llarbegin 
	\else
	\edef\tabu@temp{\ifnum\tabu@currentgrouptype=5\vtop
		\else\ifnum\tabu@currentgrouptype=12\vcenter
		\else\vbox\fi\fi}%
	\setbox\tabu@box \hbox\bgroup$\tabu@temp \bgroup
	\let\tabu@verticalspacing \tabu@verticalsp@pmb
	\fi
}
\makeatother
\newif\ifnopreprintline \nopreprintlinetrue

\title{Bisimilar Conversion of \\ Multi-valued Networks to Boolean Networks}

\author{Franck Delaplace\corref{fd}}
\ead{franck.delaplace@ibisc.univ-evry.fr}

\author{Sergiu Ivanov}
\ead{sergiu.ivanov@ibisc.univ-evry.fr}
\cortext[fd]{Corresponding author}

\address{IBISC - Paris-Saclay University, Univ. Evry}	
	\begin{abstract}
	Discrete modelling frameworks of Biological networks can be divided in two distinct
	categories: Boolean and Multi-valued. Although Multi-valued networks are more expressive 
	for qualifying the regulatory behaviours modelled by more than two values, the ability to 
	automatically convert them to Boolean network with an equivalent behaviour breaks down the 
	fundamental borders between the two approaches. Theoretically investigating the conversion process provides 
	relevant insights into bridging the gap between them. Basically, the conversion aims at finding 
	a Boolean network \emph{bisimulating} a Multi-valued one. In this article, we investigate the 
	bisimilar conversion where the Boolean integer coding is a parameter that can be freely 
	modified. Based on this analysis, we define a computational method automatically inferring 
	a bisimilar Boolean network from a given Multi-valued one.
	\end{abstract}


\maketitle

\section{Introduction}

Discrete network based modelling frameworks, seminally introduced by 
S.~Kauffman~\cite{Kauffman1969, Glass1973} and R.~Thomas~\cite{Thomas1995, Thieffry1995} for 
regulation network modelling can be divided in two distinct categories: Boolean networks and Multi-valued 
networks. In the former, the states of genes are modelled by Boolean values, with propositional logic as the
modelling framework, whereas in the latter the state is extended to the integer domain, also called Multi-valued, using 
Presburger arithmetic as a modelling framework. 
It is often admitted that Multi-valued networks provide more expressiveness for modelling gene expression behaviour by distinguishing between more than two states (\ie \textsc{off} or \textsc{on}) for specifying the regulatory activity. 
 However, the ability to automatically convert a Multi-valued network to a Boolean one with the 
 same dynamical behaviour weakens this distinction from an analytical standpoint since the analysis 
 of the dynamics can be performed on the Boolean network directly.
 
More generally, the Boolean conversion of a Multi-valued network offers the opportunity to bridge the gap between the two modelling formalisms that enables to inherit, adapt and extend the theoretical results defined in a framework to the other~\cite{Tonello2019}. Moreover, this allows the use of software based on propositional logic that could prove computationally more efficient than the algorithms developed for Presburger arithmetic for the same problem. In particular, a wide spectrum of problems in modelling regulatory networks by symbolic characterization of stable states can be formalized as problems of logical valuation of variables satisfying a formula in the Boolean case (the SAT problem) or finding solutions complying to a set of linear constraints for the integer case (integer linear programming, ILP). The work~\cite{Aloul2002} provides an experimental comparison of ILP and SAT solvers applied to the SAT problem.

By considering these opportunities, the issue is thus to investigate methods for converting Multi-valued networks to Boolean, while preserving the dynamical behaviour. This conversion is primarily based on an encoding of integers by Boolean profiles, establishing the equivalence between the two kinds of values. The challenge is to extend this equivalence to state transitions in order to certify the behavioural integrity.

In~\cite{Didier2011}, G.~Didier, E.~Remy and C.~Chaouya extensively study the conditions for the 
conversion of Multi-valued networks to Boolean ones using Van Ham code~\cite{VanHam1979} 
(Section~\ref{sec:code}). To overcome the potential limitation of Van Ham code restraining the 
dynamics to a sub-region of the Boolean state space, A.~Faur\'e and S.~Kaji study the conversion 
based on \emph{Summing code} (Section~\ref{sec:code}), which provides several alternative Boolean 
profiles for encoding an integer, such that the resulting Boolean dynamics is deployed on the whole 
Boolean state space~\cite{Faure2018}. Following similar motivations, E.~Tonello also studies the 
conversion based on this code~\cite{Tonello2019}.

These research works elegantly pave the formal foundation of the Multi-valued to Boolean network 
conversion. However, the results are intrinsically dependent on a specific coding, the Summing code, 
and are mainly designed for the asynchronous mode. Therefore, it appears interesting to generalise this 
approach by distinguishing the properties that purely relate to the conversion process from those 
depending on the code for highlighting the foundations of this process. 

The behavioural equivalence is formally defined by the \emph{reachability preservation} property, 
namely: whenever an integer state is reachable from another one, the equivalent Boolean state of 
the former is also reachable by the equivalent Boolean state of the latter, and conversely. 
Reachability preservation relies on the existence of a bisimulation~\cite{Sangiorgi2011} between both 
networks, parametrised by the Boolean-to-integer coding. While preserving the reachability is 
essential, it also appears desirable to extend the preservation to structural properties of the 
interactions and other properties related to equilibrium.
In this article, we  study the network conversion by regarding it as a bisimulation process applied to any Boolean coding of the integers. Based on this study, we propose an algorithm inferring the formulas of a Boolean network behaviourally equivalent to the input Multi-valued network. 
 
 After recalling the main notions of Multi-valued networks (Section~\ref{sec:network-theory}), we 
 examine the bisimulation properties between the dynamics of the networks and the admissibility 
 conditions for stating a bisimulation between a Multi-valued and Boolean networks 
 (Section~\ref{sec:bisimulation}) with regard to different codings (Section~\ref{sec:code}). Then, 
 we study the extension of the properties preserved by conversion 
 (Section~\ref{sec:preserving-properties}). Finally we define a method inferring a Boolean network 
 bisimilar to the Multi-valued one whatever the coding procedure 
 (Section~\ref{sec:algo-conversion}).

\paragraph{Notations} We use the following notations:
\begin{description}
\item[\it Set:] The \emph{complement} of a subset $E \setminus E', E' \subseteq E$ is denoted by 
$-E'$. A \emph{singleton} $\{e\}$ is denoted by its element $e$. The \emph{set of parts} of $E$ is 
noted $\mathbf{2}^E =\{ E' \mid E' \subseteq E \}$. 

\item[\it State:]
A \emph{state} $s$ is an application from variables $Y$ to a domain of values $\mathbb{D}$, \ie 
$s=\{ y_1 \mapsto d_1, \ldots,y_n \mapsto d_n\}$ and $\mathbb{D}_Y = (Y \to \mathbb{D})$ denotes 
the \emph{state space} defined on variables $Y$. The \emph{restriction/projection} of a state $s 
\in\mathbb{D}_Y $ on $W \subseteq Y$ is denoted $s_{W} \in \mathbb{D}_W$. This notation also holds 
for function on states, \ie $\dom g_W=\mathbb{D}_W$. A \emph{substitution} within a state $s$ is the replacement of 
the value of a variable of $s$ by another value, formally defined as:
$s_{[y \mapsto v]} = s \setminus \{y \mapsto s_y \} \cup \{y \mapsto v\}.$ 
 The \emph{distance} on states is defined as: 
$d(s, s')= \sum_{i=1}^{n} |s_{y_i} - s_{y_i}'|.$
 
\end{description}

\section{Multi-valued networks}
\label{sec:network-theory}
A \emph{Multi-valued network} $\tuple{g, Y}$ defined on a set of variables $Y$ is a dynamical 
system on integer states where the \emph{evolution function} $g:\Nset_Y \to \Nset_Y$ is composed of 
a collection of \emph{local} evolution functions $g=(g_1, \cdots, g_n), n=|Y|$. The evolution 
function is defined for a variable $y_i \in Y$ as follows:
\begin{equation}
 g_i(s) =\begin{cases}
 1 & \textit{if } C^{1} (s)\\
\multicolumn{2}{c}{\cdots}\\
l& \textit{if } C^{l}(s) \\
\multicolumn{2}{c}{\cdots}\\
L & \textit{if } C^{L}(s) \\
0 & \textit{otherwise}
 \end{cases}
 \label{eq:Multi-valued-def}
\end{equation}
where $C^l$ is the \emph{guard} of level $l$. The application $g_i(s)$ equals level $l$ if and only if the guard $C^l(s)$ is the first satisfied condition with respect to the reading order. 

\medskip
\paragraph{Model of dynamics} The \emph{model of dynamics} of a Multi-valued network $\tuple{g, Y}$ is formalized by a labelled transition system $\tuple{\Nset_Y, M, \evo_{g}}$ where the labels are sets of variables that determine which variables are updated jointly during a transition. 
The \emph{mode} $M \subseteq \partset{Y}$ describes the organization of the joint updates per transition. For example, in the \emph{asynchronous} mode, $\textbf{1}_Y=\{ \{y_i\} \}_{y_i \in Y}$, the state of one variable only is updated per transition and in the \emph{parallel} or \emph{synchronous} mode $\{Y\}$, all the variables are updated together. The mode is also introduced in the network specification if needed, \ie $\tuple{g, Y, M}$.

Thus, only the state of the variables in $m \in M $ can be updated by a transition $s \xevo{m}_g s'$ 
whereas the state of the other variables remains unchanged \ie $s'=\left(g_m(s) \cup s_{-m})\right).$
A transition that does not change the state, $s \xevo{m}_g s$, is called a \emph{self-loop}.
The global transition relation corresponds to the union of all transition relations labelled by the components of the mode: 
$ \evo_{g} = \bigcup_{m \in M} \xevo{m}_g.$

\medskip
Hereafter, $f : \Bset_X \to \Bset_X, \Bset=\{0,1\}$, always stands for a Boolean function, $g: \Nset_Y\to \Nset_Y$ designates a Multi-valued/integer function, whereas $Y$ always corresponds to a set of integer variables. $w \in \Bset_X$ represents a Boolean state whereas $s \in \Nset_Y$ a Multi-valued one. The Multi-valued dynamics where in which transitions modify the current level by $1$ only (\ie $ \forall s\evo s', \forall y_i \in Y: d(s_{y_i}, s_{y_i}') \leq 1$) is said \emph{unitary stepwise}.
 
\paragraph{Equilibrium}
A state $s$ is an \emph{equilibrium}, if it can be reached\footnote{$\evo^*$ denotes the reflexive 
and transitive closure of $\evo$.} infinitely once met:
\begin{equation}
 \forall s'\in \Nset_Y:s \evo^* s' \implies s' \evo^* s.
\end{equation}
An \emph{attractor} is a set of equilibria that are mutually reachable and a \emph{stable state} is an attractor of cardinality~$1$.

\begin{figure}
	\begin{center}
		\begin{tabular}{c c}
		\begin{tabular}{c}	
		 $g=
			\begin{cases}
			x = &
			\begin{cases}
			1 & y\geq 1 \\
			0 & \textit{otherwise} \\
		 \end{cases}
			\\
			\\
			y = &
			\begin{cases}
			3 & x=1\land y\geq 2 \\
			2 & (x=1\land y=1)\lor (x=0\land y=3) \\
			1 & (x=1\land y=0)\lor (x=0\land y=2) \\
			0 & \textit{otherwise}
			\end{cases}
	 \end{cases}$	
	 \\
	 \begin{tikzpicture}[baseline=3ex]
			\GraphInit[vstyle=Normal]
			\SetGraphUnit{3}
			\Vertex[L=$x$] {X}
			\EA[L=$y$](X){Y}
			\SetUpEdge[style={->,bend left, thick, post, sloped}]
			\tikzstyle{LabelStyle}=[fill=white,sloped]
			\Edge[label=$+$](X)(Y)
			\Edge[label=$+$](Y)(X)
	 	 \Loop[label=$+$,dist=1.5cm,dir=EA](Y)
			\end{tikzpicture}
			\end{tabular}
			
			&
			\begin{tikzpicture}[baseline=15ex, xscale=2, yscale=1.5]
			\GraphInit[vstyle=Normal]
			\tikzset{VertexStyle/.append style={font=\tiny, draw, shape=rectangle, line width = 
			0.5, inner sep = 2.5pt, outer sep = 0.5pt, minimum size = 0}}
			\tikzset{EdgeStyle/.style={post}}
			\tikzset{LabelStyle/.style={font=\footnotesize, sloped}} 
			\definecolor{colorof00}{gray}{0.85} \tikzset{VertexStyle/.append 
			style={fill=colorof00}}
			\Vertex[x=1., y=1.]{00}
			\tikzset{VertexStyle/.append style={fill=white}}
			\definecolor{colorof13}{gray}{0.6} \tikzset{VertexStyle/.append 
			style={fill=colorof13}}
			\Vertex[x=2., y=4.]{13}
			\tikzset{VertexStyle/.append style={fill=white}}
			\Vertex[x=1., y=2.]{01}
			\Vertex[x=1., y=3.]{02}
			\Vertex[x=1., y=4.]{03}
			\Vertex[x=2., y=1.]{10}
			\Vertex[x=2., y=2.]{11}
			\Vertex[x=2., y=3.]{12}
			\Edge[label={$y$}](01)(00)
			\Edge[label={$x$}](01)(11)
			\Edge[label={$y$}](02)(01)
			\Edge[label={$x$}](02)(12)
			\Edge[label={$y$}](03)(02)
			\Edge[label={$x$}](03)(13)
			\Edge[label={$x$}](10)(00)
			\Edge[label={$y$}](10)(11)
			\Edge[label={$y$}](11)(12)
			\Edge[label={$y$}](12)(13)
			\end{tikzpicture}
		\end{tabular}
	\end{center}
	\caption{A Multi-valued network with the interaction graph (below) and the asynchronous 
	dynamics (right), with the self-loops removed.}
	\label{fig:Multi-valued-network-ex}
\end{figure}
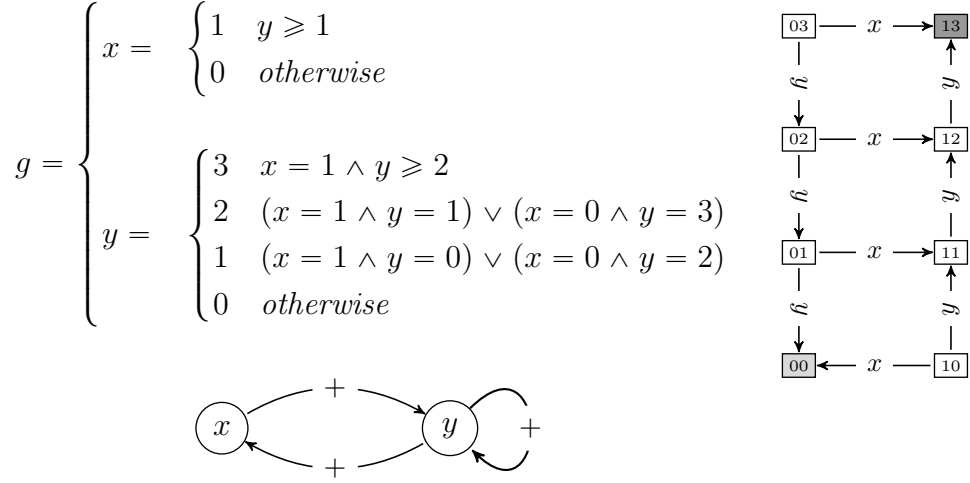

Figure~\ref{fig:Multi-valued-network-ex} shows an example of a Multi-valued network and the resulting dynamics for the asynchronous mode with two stable states that are respectively $13$ and $00$. 
\paragraph{Interaction graph} An \emph{interaction graph} $\tuple{Y, \arc}$ portrays the interdependencies of the variables in the network $\tuple{g, Y}$. An interaction $y_i \arc y_j$ exists whenever changing the value of $y_i$ may lead to a change in the value of $y_j$:
\begin{equation}
 y_i \arc y_j \eqdef \exists s, s' \in \Nset_Y: s_{y_i} \neq s'_{y_i} \land s_{-y_i}=s'_{-y_i} \land g_j(s)\neq g_j(s').
 \label{eq:interaction}
\end{equation}

The \emph{signed interaction graph} $\tuple{Y, \arc, \sigma}$ refines the nature of the interactions by 
 signing the arcs with $\sigma: (\arc) \to \{-1, 0, 1\}$ to represent a monotone relation between 
 the source and target variables of the interaction~(\ref{eq:signed-interaction}); either increasing 
 (label $1$, denoted $'+'$), or decreasing (label $-1$, denoted $'-'$), or neither (label $0$, denoted $'\pm'$), and formally defined as:
\begin{equation}
\begin{split}
 y_i \sarc{+} y_j &\eqdef y_i \arc y_j \land \\
 \forall s, s' &\in \Nset_Y: s_{y_i} \leq s'_{y_i} \land s_{-y_i} = s'_{-y_i} \implies g_{j}(s)\leq g_{j}(s') \\
 y_i \sarc{-} y_j & \eqdef y_i \arc y_j \land \\
 	\forall s, s'& \in \Nset_Y: s_{y_i} \leq s'_{y_i} \land s_{-y_i} = s'_{-y_i} \implies g_{j}(s)\geq g_{j}(s')
\end{split}
\label{eq:signed-interaction}
\end{equation}

\section{Network bisimulation}
\label{sec:bisimulation}
\begin{figure}[h!]
 	\centering
 	\begin{subfigure}[b]{.4\textwidth}
 		\centering
 		\begin{tikzpicture}[node distance=8mm and 15mm]
 		\node (w) {$w$\strut};
 		\node[below=of w] (w') {$w'$\strut};
 		\draw[->] (w) -- node[midway, auto] (m) {\small $m$} (w');
 		
 		\node[right=of w] (s) {$ \psi(w)$\strut};
 		\node[below=of s] (s') {$ \psi(w')$\strut};
 		\draw[->] (s) -- node[midway, auto] (n) {\small $\mu(m)$} (s');
 		
 		\draw[->, densely dashed] (w) -- (s);
 		\draw[->, densely dashed] (w') -- (s');
 		\draw[->, densely dashed, shorten >=2mm] (m) -- (n);
 		\end{tikzpicture}
 		\caption{Forward simulation}
 	\end{subfigure}
 	\begin{subfigure}[b]{.4\textwidth}
 		\centering
 		\begin{tikzpicture}[node distance=8mm and 15mm]
 		\node (s) {$s$\strut};
 		\node[below=of s] (s') {$s'$\strut};
 		\draw[->] (s) -- node[midway, auto] (n) {\small $n$} (s');
 		
 		\node[right=of s] (w) {$\forall w\in \psi^{-1}(s)$\strut};
 		\node[below=of w] (w') {$\exists w'\in \psi^{-1}(s')$\strut};
 		\draw[->] (w) -- node[midway, auto] (m) {\small $\exists m\in \mu^{-1}(n)$} (w');
 		\draw[->, densely dashed] (s) -- (w);
 		\draw[->, densely dashed] (s') -- (w');
 		\draw[->, densely dashed, shorten >=2mm] (n) -- (m);
 		\end{tikzpicture}
 		\caption{Backward simulation}
 	\end{subfigure}
 	\caption{Illustration of a bisimulation $(\psi, \mu)$ between a
 		Boolean and a Multi-valued network. $ \psi^{-1}(s)$ and
 		$\mu^{-1}(n)$ denote the preimages of $s$ and $n$ under $ \psi$
 		and $\mu$ respectively.}
 	\label{fig:bisim}
 \end{figure}
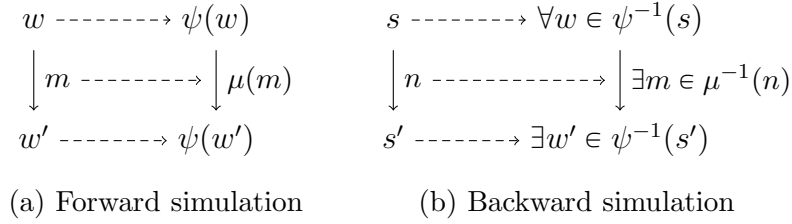

By definition~\cite{Sangiorgi2011}, bisimulation between the dynamics of networks preserves the reachability, thereby maintaining the trajectories and the attractors in both ways.
Definition~\ref{def:bisimulation} illustrated in Figure~\ref{fig:bisim} formally defines functional bisimulation, which depends on a partial function $ \psi: \Bset_X \to \Nset_Y$ decoding a Boolean state to an integer state. 

 \begin{definition}
 	\label{def:bisimulation}
 	Given a Boolean network $B=\langle f, X, M_X\rangle $ and a
 	Multi-valued network $N=\langle g, Y, M_Y\rangle$, a pair of
 	functions $(\psi, \mu)$, with $ \psi:\Bset_X\to \Nset_Y$ a partial function and $\mu:M_X\to M_Y$ a total function, form a bisimulation if and only if
 	the following properties hold:
 	\begin{enumerate}
 		\item\label{def:bisim-ftog} (forward simulation) for any two Boolean states
 		$w, w'\in
 		\dom \psi$ and $m\in M_X$, $w\xevo{m}_f w'$ implies
 		$ \psi(w)\xevo{\mu(m)}_g \psi(w')$:
 		\[
 		\forall w, w'\in \dom \psi, \, \forall m\in M_X : w\xevo{m}_f w' \implies \psi(w)\xevo{\mu(m)}_g \psi(w');
 		\]
 		
 		\item\label{def:bisim-gtof} (backward simulation) for any two Multi-valued states
 		$s, s'\in
 		\Nset_Y$, for any $w\in \Bset_X$ such that $ \psi(w)=s$, and
 		for any $n\in M_Y$, $s\xevo{n}_g s'$ implies that there exists a
 		$w'\in \Bset_X$ and an $m\in M_X$ such that $ \psi(w') = s'$, 
 		$\mu(m) = n$ and $w\xevo{m}_f w'$:
 		\begin{multline*}
 		\forall s, s'\in \Nset_Y, \, \forall w\in \Bset_X, \, \forall n\in M_Y :
 		\psi(w)=s \land s\xevo{n}_g s' \implies \\
 		\big(\exists w'\in \Bset_X, \, \exists m\in M_X : \mu(m) = n\land \psi(w') = s' \land w\xevo{m}_f w' \big).
 		\end{multline*}
 	\end{enumerate}
 \end{definition}
 
Two networks $B$ and $N$ complying to Definition~\ref{def:bisimulation} with respect to $ \psi$ are 
said \emph{bisimilar}, noted $B \sim_ \psi N$.
Although, (\ref{def:bisimulation}.\ref{def:bisim-gtof}) and 
(\ref{def:bisimulation}.\ref{def:bisim-ftog}) are similar in their definition, it is worth noticing 
that they however differ in the following point: all the transitions on the integer state space should 
fulfill (\ref{def:bisimulation}.\ref{def:bisim-gtof}) whereas only the transitions defined on the 
domain of $ \psi$, $\dom \psi$, should comply to (\ref{def:bisimulation}.\ref{def:bisim-ftog}). 
$\dom \psi$ circumscribes the \emph{admissible region} \cite{Faure2018, Tonello2019}, where each 
Boolean state encodes an integer state and each transition is bisimilar to a Multi-valued one. 
Hence, no transitions from a state located in the admissible region can escape from this region, 
thus avoiding aberrant cases exemplified in \cite{Tonello2019}.
 From (\ref{def:bisimulation}.\ref{def:bisim-gtof}), we deduce that $ \psi$ is a surjective partial function defined on $\Bset_X$ but it is not necessary injective and thus not bijective. Hence, the preimage of an integer state is a set:
$ \psi^{-1}(s)=\{w \in \Bset_X \mid \psi(w)=s\}.$

The issue is to determine the conditions on a Boolean network enabling a bisimulation with a Multi-valued network. These conditions depend on a general relation between the integer and Boolean function including the mode.

\subsection{From global to local bisimulation discovery}
\label{subsec:global-to-local-bisimulation}

Integer states are coded by the Boolean states in which the Boolean variables storing the code constitute the \emph{support} of the integer variables. The \emph{support function} associates each subset of integer variables to its support: $\sprt{\cdot}: \partset{Y} \to \partset{X}$. This function has the following properties: 1) the Boolean variables are exactly the supports of the integer variables, 2) the supports are pairwise disjoint, and 3) they are modular in the sense that the union of the supports is the support of the union of the integer variables:
 \begin{equation}
 \begin{tabu} to 0.8\textwidth {l X[5,l,$$] X[1,c] }
 1) &X = \sprt{Y}& and, \\
 2) & \forall y_i, y_j \in Y: y_i \neq y_j \iff \sprt{y_i} \cap \sprt{y_j} = \emptyset & and, \\
 3)\quad &	 \forall Y', Y" \subseteq Y: \sprt{Y' \cup Y"}=\sprt{Y'} \cup \sprt{Y"}.\\
 \end{tabu}
 \label{eq:support}
 \end{equation} 
 For example, the state $s_{(y_1,y_2)}=(0,1)$ is encoded by $w=(00,01)$ by using the classical binary code or the Gray code. The variables of the Boolean network will be therefore the variables supporting the Boolean code of the integer variables of $Y$ ($X=\sprt{Y}$).
The states of said Boolean variables are respectively: $w_{\sprt{y_1}_1}=0,w_{\sprt{y_1}_2}=0, w_{\sprt{y_2}_1}=0, w_{\sprt{y_2}_2}=1$.  Note that there are two kinds of indices: one for the multivalued variables, and the other for the Boolean variables of the corresponding supports.

We consider henceforth that $\psi$ fits all supports, \ie $\psi \in \bigcup_{W \subseteq Y} 
\left(\Bset_{\sprt{W}} \to \Nset_{W}\right)$.
 The function $\psi$ transforms the Boolean state of the support into an integer state in a modular manner, by decoding distinct sub-parts of a Boolean state separately, so that the decoding of the whole integer state is the union of the local decoding results:
\begin{equation}
 \forall W \subseteq Y, \forall w \in \Bset_{\sprt{Y}}: \psi(w_{\sprt{W}})= \psi(w)_{W}.
 \label{eq:support-fun}
\end{equation}
From (\ref{eq:support},\ref{eq:support-fun}) we deduce the following
relation on two disjoint sets of variables representing the modularity of the decoding:
\begin{equation*}
 \psi(w_{\sprt{W \cup W'}})= \psi(w_{\sprt{W}}) \cup \psi(w_{\sprt{W'}})=\psi(w)_{W} \cup 
 \psi(w)_{W'}, W \cap W'=\emptyset, W,W' \subseteq Y.
\end{equation*}

Moreover, the  mode of the converted
Boolean network must be compatible with the modularity of the
coding. A mode is \emph{local-to-support} when the parallel updates of the
Boolean variables operate inside supports only, namely
$M$ is local-to-support if and only if:
$\forall m \in M, \exists y_i \in Y: m \subseteq \sprt{y_i}.$
The asynchronous mode is always local-to-support and the parallel
local-to-support mode is the gathering of the supports:
$\{\sprt{y_i}\}_{y_i \in Y}.$

 \medskip
 Within the framework, inferring a Boolean network bisimilar to a multivalued one is reduced to the discovery of a Boolean network in bisimulation with a \emph{local multivalued network} $\tuple{g_i, Y}$ where only the state of a single variable evolves. A Boolean network in bisimulation with the entire multivalued network results from the union of Boolean networks in bisimulation with local multivalued networks (Proposition~\ref{prop:localtoglobal}). Hence, for each $g_i$, we focus on the discovery of the appropriate evolution function of the support $f_{\sprt{y_i}}$ and the determination of the admissible modes for enabling the bisimulation. 

 \begin{proposition}
 \label{prop:local-to-global-bisimulation}
 Consider the multivalued network $N = \tuple{g, Y, \textbf{1}_Y}$
 and the family $(B_i)_{y_i\in Y}$ of Boolean networks over the
 supports of the variables in $Y$:
 $B_i = \tuple{f_{\sprt{y_i}}, X, M_{y_i}}$,
 $M_{y_i} \subseteq \partset{\sprt{y_i}}$. Then the following
 holds:
 \[
 \big(\,\forall y_i \in Y :
 \tuple{f_{\sprt{y_i}}, X, M_{y_i}} \sim \tuple{g_i, Y, \textbf{1}_{y_i}}\,\big)
 \implies
 \tuple{f, X, \bigcup_{y_i \in Y} M_{y_i}} \sim \tuple{g, Y, \textbf{1}_Y}.
 \]
 where $f=(f_{\sprt{y_i}})_{y_i \in Y}$ and $g=(g_i)_{y_i \in Y}$
 are the global evolution functions collecting their respective
 local evolution functions.
\label{prop:localtoglobal}
\end{proposition}

According to the definition of a transition (Section~\ref{sec:network-theory}) and
to Proposition~\ref{prop:local-to-global-bisimulation}, to establish a
bisimulation relation between a multivalued transition function $f$
and a Boolean transition function $g$, it suffices that the
\emph{local} integer evolution function applied to the decoding
$g_i\circ \psi(w)$ coincide with the Boolean evolution function
${\psi}\circ f_{\sprt{y_i}}(w)$ or, more generally, with the
evolution  taken under a local-to-support mode:
\begin{equation}
\centering
\begin{tabu} to 0.9\textwidth { c @{\;=\;} X[1,l,$] X[1.6,l]}
\psi \circ f & g \circ \psi & {Global function}\\
\psi(f_{m}(w) \cup w_{\sprt{y_i} \setminus m}) & g_i \circ \psi(w), m \subseteq \sprt{y_i} &{Local-to-support mode}
\end{tabu}
\label{eq:main-fg}
\end{equation}

If $ \psi$ is a bijective function then (\ref{eq:main-fg}) is expressed as
$f = \psi^{-1}\circ g \circ \psi$,
which corresponds to the conjugated evolution function defined in~\cite{Didier2011}.

Theorem~\ref{thm:bisimulation-psi.f=g.psi-equivalence-local-to-support} shows that 
Property~(\ref{eq:main-fg}) is necessary and sufficient to ascertain that a multivalued network 
bisimulates a Boolean network with a local-to-support mode.

\begin{theorem}
	Let $N=\tuple{g_i, Y, \textbf{1}_{y_i}}$ be a multivalued network,
	$B=\tuple{f_{\sprt{y_i}}, \sprt{Y}, M}$ a Boolean
	network with $M$  a local-to-support mode, and $ \psi:\Bset_{\sprt{y_i}} \to 
	\Nset_{y_i}$, a 
	surjective
	function,
	 Property~\ref{eq:main-fg} is met between the evolution
	functions of $B$ and $N$ if and only if $B \sim_{ \psi} N$.
	\label{thm:bisimulation-psi.f=g.psi-equivalence-local-to-support}
\end{theorem}

\subsection{Bisimulation admisibility}
\label{subsec:bisimulation-admissibility}
\newcommand{\adm}[0]{\operatorname{adm}}

Theorem~\ref{thm:bisimulation-psi.f=g.psi-equivalence-local-to-support} states the equivalence 
between bisimilarity and Property~\ref{eq:main-fg} for local-to-support modes, including the parallel mode updating all the variables of the support.
In this section, we extend this result to a larger family of modes that are \emph{admissible} with respect to the parallel mode. Informally, a modality $m$ is $m_0$-admissible if, for any Boolean state $w$, running $f$ on $w$ under $m$ or under $m_0$ yields
(possibly different) states belonging to the same preimage under
$ \psi$. Definitions~\ref{def:adm-modality} and~\ref{def:adm-mode}
detail this compatibility formally.
 Both definitions assume the Boolean network $B=\tuple{f, \sprt{Y}, M_0}$ operating in mode
$M_0$ and the Multi-valued network $N=\tuple{g_i, Y, \textbf{1}_{y_i}}$.


\begin{definition}\label{def:adm-modality}
A mode component $m\in M$ is {\em $m_0$-admissible with
 respect to the functional bisimulation $B_0\sim_{ \psi} N$} denoted by $\adm_{ \psi}(m, m_0)$, if the following holds:
 \[
 \forall w\in \Bset_{\sprt{Y}} : \psi(f_{m_0}(w)\cup w_{-m_0}) = \psi(f_m(w)\cup w_{-m}).
 \]
\end{definition}

Notice that it follows directly from the definition that admissibility
is an equivalence relation on modalities. Admissibility is defined on the modalities composing a mode and can be lifted from modalities to modes in a natural way:

\begin{definition}\label{def:adm-mode}
 A mode $M$ is {\em $M_0$-admissible with respect to the functional
 bisimulation $B\sim_{ \psi} N$}, denoted by $\adm_{ \psi}(M, 
 M_0)$, iff the following conditions hold:
 \begin{enumerate}
 \item $\forall m_0\in M_0, \, \exists m\in M \, :\, \adm_{ \psi}(m, m_0)$;
 \item $\forall m\in M, \, \exists m_0\in M_0 \, :\, \adm_{ \psi}(m, m_0)$.
 \end{enumerate}
\end{definition}

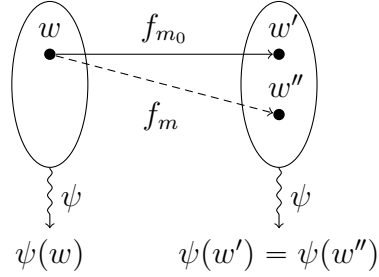
\begin{figure}[h]
	\centering
	\begin{tikzpicture}[node distance=8mm and 20mm]
	\tikzstyle preimage=[draw, ellipse, minimum height=22mm, minimum width=10mm]
	\tikzstyle state=[fill, circle, inner sep=0, minimum size=1.5mm]
	\newcommand{\dotspread}{4mm}
	
	\node[preimage] (pre1) {};
	\node[preimage, right=of pre1] (pre2) {};
	
	\node[state, yshift=\dotspread, label=90:$w$] (w) at (pre1) {};
	\node[state, yshift=\dotspread, label=90:$\;w'$] (w') at (pre2) {};
	\node[state, yshift=-\dotspread, label=90:$\;\, w''$] (w'') at (pre2) {};
	
	\draw[->] (w) -- node[midway, auto] {$f_{m_0}$} (w');
	\draw[->, densely dashed] (w) -- node[pos=.62, auto, swap] {$f_m$} (w'');
	
	\node[below=of pre1] (alpha w) {$ \psi(w)$};
	\node[below=of pre2] (alpha w') {$ \psi(w') = \psi(w'')$};
	
	\tikzstyle alpha=[->, decorate, decoration={snake, 
		amplitude=.3mm, segment length=2mm, post length=1mm}]
	\draw[alpha] (pre1) -- node[midway, auto] {$ \psi$} (alpha w);
	\draw[alpha] (pre2) -- node[midway, auto] {$ \psi$} (alpha w');
	\end{tikzpicture}
	\caption{Illustration of $m_0$-admissibility $\adm_{ \psi}(m, m_0)$ of a modality $m$ with
		respect to the bisimulation $B\sim_{ \psi} N$.}
	\label{fig:adminissibility}
\end{figure}

According to the definition, a mode $M$ is $M_0$-admissible if, for
every modality $m\in M$, there exists a modality $m_0\in M_0$ such that
$m$ is $m_0$-admissible. Note that this requirement
does not imply the existence of a bijection between $M$ and
$M_0$: two functions $M_0\to M$ and $M\to M_0$ are indeed required
by, respectively, clauses (1) and (2) of Definition~\ref{def:adm-mode}, but they may not be the inverse of each other.

\begin{lemma}\label{lem:adm-mode-eq}
 The relation of admissibility with respect to the functional
 bisimulation $B\sim_{ \psi} N$, defined on all possible
 modes of $B$, is an equivalence relation.
\end{lemma}

 It turns out that switching update modes {\em within} an admissibility
class of modes preserves bisimulation.

\begin{theorem}\label{thm:adm-bisimulation}
 Given the functional bisimulation $B\sim_{ \psi} N$ between the
 Boolean network $B=\tuple{f, \sprt{Y}, M_0}$ and the Multi-valued
 network $N=\tuple{g_i, Y, \textbf{1}_{y_i}}$, any Boolean network
 $B'=\tuple{f, \sprt{Y}, M}$ with $\adm_{ \psi}(M, M_0)$
 functionally bisimulates $N$ as well:
 \[
 \forall M\subseteq \mathbf{2}^{\sprt{Y}} \, :\, B\sim_{ \psi} N \land
 \adm_{ \psi}(M, M_0) \implies B'\sim_{ \psi} N.
 \]
\end{theorem}

This result allows us to prove the bisimilarity of a network with another mode providing that 
Property~(\ref{eq:main-fg}) holds and the mode is admissible. 

\begin{corollary}
\label{cor:admissible-mode-and-psi.f=g.psi}
 Let
 $B=\tuple{f, \sprt{Y}, M}$, $B'=\tuple{f, \sprt{Y}, M'}$ be two Boolean networks,
 $ \psi:\Bset_{\sprt{Y}} \to \Nset_Y$ a surjective function, and
  $N=\tuple{g_i, Y, \textbf{1}_{y_i}}$  a Multi-valued network.
  If  Property~\ref{eq:main-fg} holds for $B$, $N$, and $\psi$, and if $\adm_{\psi}(M, M')$, then $B' \sim_{ \psi} N.$
\begin{proof}
If 	 Property~\ref{eq:main-fg} holds then we deduce that $B \sim_{ \psi} N$ from 
Theorem~\ref{thm:bisimulation-psi.f=g.psi-equivalence-local-to-support}. As $B \sim_{\psi} N$ and 
$M$ is an admissible mode we conclude from Theorem~\ref{thm:adm-bisimulation} that $B' \sim_{ \psi} 
N$.
\end{proof}
\end{corollary}

\section{Boolean coding}
\label{sec:code}
The coding procedure characterizes a function $\psi$ mapping a Boolean profile to an integer. We 
study two fundamental codes that are suitable for asynchronous Boolean dynamics: the \emph{Summing 
code} and the \emph{Gray code}. Table~\ref{tab:code} shows both codings for encoding levels ranging from 
$0$ to $3$. 

\paragraph{Summing code} For the Summing code, the integer corresponding to a Boolean state $w$ is the sum of the states of the Boolean support variables:
$$\psi(w_{\sprt{y_i}})=\sum_{\sprt{y_i}_k \in \sprt{y_i}} w_{\sprt{y_i}_k}.$$ 
 The size of the support is linear in the maximal level, $|\sprt{y_i}|=L$, and different encodings are
 possible for the same integer. The number of different codes for an integer $0 \leq l \leq L$ is 
 $\binom{L}{l}$. Van Ham code~\cite{VanHam1979} is a sub-case of the Summing code in which the unitary 
 stepwise evolution restricts the filling of $1$ from left to right. This code is emphasized in 
 bold in Table~\ref{tab:code}.
 
 \paragraph{Gray code} The Gray code associates Boolean states differing in only one 
 position to consecutive integers. The coding function is bijective and constructs the integer value from a Boolean state by 
 first transforming a Gray code profile into its equivalent binary code and then by computing the 
 integer from this coding\footnote{$\oplus$ is the exclusive \textsc{or}, \textsc{xor}.}:
 $$\psi(w_{\sprt{y_i}})= \sum_{k=1}^{|\sprt{y_i}|} 2^{|\sprt{y_i}| - k }.\bigoplus_{j=1}^{k} w_{\sprt{y_i}_{j}}.$$
 The support size is logarithmic in the maximal level: $|\sprt{y_i}|=\lceil \log_2(L+1) \rceil.$
 
\begin{table}[ht]
\centering
\begin{tabular}{c}
\begin{tikzpicture}
\matrix(sum)[row 1/.style={nodes={fill=gray!20, align=center, text width=6.em, font=\bfseries}}, 
matrix of math nodes, row sep=0.5em, column sep=.85mm]
{
	{0} & {1} & {2} & {3} \\
};
\end{tikzpicture}
\\
\begin{tikzpicture}
\matrix(sum)[ 
matrix of math nodes, row sep=1em, column sep=3em]
{
 & (0, 1, 0) & \bf (1, 1, 0) & \\
 \bf (0, 0, 0) & \bf (1, 0, 0) & (0, 1, 1) & \bf (1, 1, 1)\\
 & (0, 0, 1) & (1, 0, 1) & \\
};
\foreach \i in {1,..., 3}
\draw[thick] (sum-2-1) -- (sum-\i-2);
\draw[thick] (sum-1-2) -- (sum-1-3);
\draw[thick] (sum-1-2) -- (sum-2-3);
\draw[thick] (sum-2-2) -- (sum-1-3);
\draw[thick] (sum-2-2) -- (sum-3-3);
\draw[thick] (sum-3-2) -- (sum-2-3);
\draw[thick] (sum-3-2) -- (sum-3-3);

\foreach \i in {1,..., 3}
\draw[thick] (sum-\i-3) -- (sum-2-4);
\end{tikzpicture} 
\\
\textsc{- Summing code -}
\\
\begin{tikzpicture}
\matrix(graycode)[
matrix of math nodes, row sep=0.5em, column sep=4em]
{
	(0, 0) & (0,1) & (1, 1) & (0, 1) \\
};
\draw[thick] (graycode-1-1) -- (graycode-1-2);
\draw[thick] (graycode-1-2) -- (graycode-1-3);
\draw[thick] (graycode-1-3) -- (graycode-1-4);
\end{tikzpicture}
\\ 
\textsc{- Gray code -} 
\end{tabular}

\caption{Example of codes for levels ranging from $0$ to $3$. The states correspond to the variable 
profiles $(\sprt{y_i}_1, \sprt{y_i}_2, \sprt{y_i}_3)$. The links connect codes differing by $1$,
and the codes in bold correspond to Van Ham sequence. }
\label{tab:code}
\end{table}
 
\medskip
The Summing code is defined on the whole Boolean state space (\ie $\dom \psi = \Bset_{\sprt{Y}}$). The Gray 
code can be also defined on the whole Boolean space when the maximal number of levels is $L=2^k-1$.
Van Ham code, on the other hand, never covers the entire Boolean space, except when the maximal level is $1$. All these codings associate the integer $0$ 
 to the $0$ Boolean profile. Furthermore, they all fulfil the \emph{neighbourhood preserving} property 
(\ref{eq:neighbourhood-preserving}) defined in \cite{Didier2011} and stressing that the distance of
$1$ between two integer states should map to a distance of 1 between the corresponding Boolean states, and~conversely: 
\begin{equation}
\begin{split}
&\forall s,s' \in \Nset_Y: d(s,s')=1 \implies \exists w \in \psi^{-1}(s),\exists w' \in 
\psi^{-1}(s'): d(w,w')=1 \;\land \\
&\forall w,w' \in \dom \psi: d(w,w')=1 \implies d(\psi(w),\psi(w')) =1 
.
\end{split}
\label{eq:neighbourhood-preserving}
\end{equation}

These codes are individual representatives of families of linear and $\log$-size codes which can be obtained by a 
permutation $\pi$ on the integer states, \ie $\psi'= \pi \circ \psi$. This permutation may notably relax the 
neighbourhood preserving property.
In the literature, the study of the Multi-valued-to-Boolean network conversion has been carried out extensively for 
the Summing and Van Ham codes~\cite{Didier2011,Faure2018, Tonello2019}. Although the Gray code is bijective 
and provides the most compact binary representation of integers, it has never been studied for the 
conversion purposes according to our knowledge. 

\section{Extensions of property preservation}
\label{sec:preserving-properties}
Although bisimulation preserves the essential property of reachability, it appears desirable to preserve additional properties for performing an accurate analysis of dynamics on the Boolean network directly. These additional properties pertain to the nature of equilibria and the interaction graph. 

\subsection{Preservation of stability of the equilibria}
\label{sec:nature-equilibria-preserving}
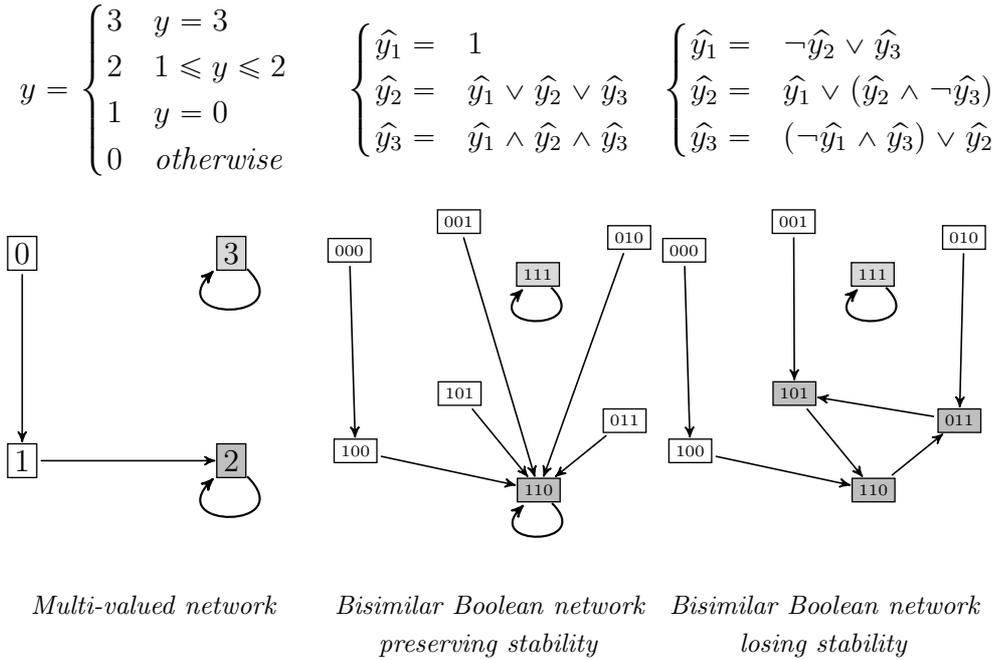
\begin{figure}[ht]
	\begin{center}
		\begin{tabu}to \textwidth { X[1,c] @{~} X[1,c] @{~} X[1,c]}
			$ y=\begin{cases}
			3 & y=3 \\
			2 & 1\leq y\leq 2 \\
			1 & y=0 \\
			0 & \textit{otherwise}
			\end{cases} $
			&
			$
		 \begin{cases}
				\sprt{y_1}= & 1 \\
				\sprt{y_2}= & \sprt{y_1}\lor \sprt{y_2}\lor \sprt{y_3} \\
				\sprt{y_3}= & \sprt{y_1}\land \sprt{y_2}\land \sprt{y_3} \\
			\end{cases}
			$
				&
			$\begin{cases}
			\sprt{y_1}= & \neg \sprt{y_2}\lor \sprt{y_3} \\
			\sprt{y_2}= & \sprt{y_1}\lor \left(\sprt{y_2}\land \neg \sprt{y_3}\right) \\
			\sprt{y_3}= & \left(\neg \sprt{y_1}\land \sprt{y_3}\right)\lor \sprt{y_2}
			\end{cases} $
			\\
			\begin{tikzpicture}[baseline={([yshift=10pt] current bounding box.north)},xscale=2.75,yscale=2.75]
			\GraphInit[vstyle=Normal]
			\tikzset{VertexStyle/.append style={font=\normalsize,draw,shape=rectangle,line width = 0.5,inner sep = 2.5pt,outer sep = 0.5pt,minimum size = 0}}
			\tikzset{EdgeStyle/.style={post}}
			\tikzset{LabelStyle/.style={font=\footnotesize, sloped}} 
			\definecolor{colorof2}{gray}{0.75} \tikzset{VertexStyle/.append style={fill=colorof2}}
			\Vertex[x=1,y=0]{2}
			\tikzset{VertexStyle/.append style={fill=white}}
			\definecolor{colorof3}{gray}{0.85} \tikzset{VertexStyle/.append style={fill=colorof3}}
			\Vertex[x=1,y=1]{3}
			\tikzset{VertexStyle/.append style={fill=white}}
			\Vertex[x=0,y=1]{0}
			\Vertex[x=0,y=0]{1}
			\Edge[](0)(1)
			\Edge[](1)(2)
			\Loop[ dist=2ex, dir=SO](2)
			\Loop[ dist=2ex, dir=SO](3)
			\end{tikzpicture}
			&
	\begin{tikzpicture}[baseline={([yshift=0pt] current bounding box.north)},xscale=1.,yscale=1.]
	\GraphInit[vstyle=Normal]
	\tikzset{VertexStyle/.append style={font=\tiny,draw,shape=rectangle,line width = 0.5,inner sep = 2.5pt,outer sep = 0.5pt,minimum size = 0}}
	\tikzset{EdgeStyle/.style={post}}
	\tikzset{LabelStyle/.style={font=\footnotesize, sloped}} 
	\definecolor{colorof110}{gray}{0.75} \tikzset{VertexStyle/.append style={fill=colorof110}}
	\Vertex[x=4.88,y=2.65]{110}
	\tikzset{VertexStyle/.append style={fill=white}}
	\definecolor{colorof111}{gray}{0.85} \tikzset{VertexStyle/.append style={fill=colorof111}}
	\Vertex[x=4.87,y=5.51]{111}
	\tikzset{VertexStyle/.append style={fill=white}}
	\Vertex[x=2.39,y=5.83]{000}
	\Vertex[x=3.83,y=6.21]{001}
	\Vertex[x=6.07,y=6.]{010}
	\Vertex[x=6.01,y=3.57]{011}
	\Vertex[x=2.47,y=3.17]{100}
	\Vertex[x=3.84,y=3.93]{101}
	\Edge[](000)(100)
	\Edge[](001)(110)
	\Edge[](010)(110)
	\Edge[](011)(110)
	\Edge[](100)(110)
	\Edge[](101)(110)
	\Loop[ dist=2em, dir=SO](110)
	\Loop[ dist=2em, dir=SO](111)
	\end{tikzpicture}
	& 
		\begin{tikzpicture}[baseline={(current bounding box.north)},xscale=1.,yscale=1.]
	\GraphInit[vstyle=Normal]
	\tikzset{VertexStyle/.append style={font=\tiny,draw,shape=rectangle,line width = 0.5,inner sep = 2.5pt,outer sep = 0.5pt,minimum size = 0}}
	\tikzset{EdgeStyle/.style={post}}
	\tikzset{LabelStyle/.style={font=\footnotesize, sloped}} 
	\definecolor{colorof011}{gray}{0.75} \tikzset{VertexStyle/.append style={fill=colorof011}}
	\Vertex[x=6.01,y=3.57]{011}
	\tikzset{VertexStyle/.append style={fill=white}}
	\definecolor{colorof101}{gray}{0.75} \tikzset{VertexStyle/.append style={fill=colorof101}}
	\Vertex[x=3.84,y=3.93]{101}
	\tikzset{VertexStyle/.append style={fill=white}}
	\definecolor{colorof110}{gray}{0.75} \tikzset{VertexStyle/.append style={fill=colorof110}}
	\Vertex[x=4.88,y=2.65]{110}
	\tikzset{VertexStyle/.append style={fill=white}}
	\definecolor{colorof111}{gray}{0.85} \tikzset{VertexStyle/.append style={fill=colorof111}}
	\Vertex[x=4.87,y=5.51]{111}
	\tikzset{VertexStyle/.append style={fill=white}}
	\Vertex[x=2.39,y=5.83]{000}
	\Vertex[x=3.83,y=6.21]{001}
	\Vertex[x=6.07,y=6.]{010}
	\Vertex[x=2.47,y=3.17]{100}
	\Edge[](000)(100)
	\Edge[](001)(101)
	\Edge[](010)(011)
	\Edge[](011)(101)
	\Edge[](100)(110)
	\Edge[](101)(110)
	\Edge[](110)(011)
	\Loop[ dist=2em, dir=SO](111)
	\end{tikzpicture}
			\\	\\
			\textit{\footnotesize Multi-valued network} & \textit{\footnotesize Bisimilar Boolean network preserving stability} & \textit{\footnotesize Bisimilar Boolean network losing stability}
		\end{tabu}
	\end{center}
	\caption{Stability loss during bisimulation -- synchronous mode.}
	\label{fig:equilibria-different-nature}
\end{figure}

By definition of the bisimulation, the equilibria of a Multi-valued network match with the equilibria of a bisimilar Boolean network, and 
conversely. However, when some equilibria are stable states, their nature may differ:
a stable state of the Multi-valued network can be represented by a cyclic attractor
over Boolean profiles, all coding for the same integer
(Figure~\ref{fig:equilibria-different-nature}).
 Nevertheless, any cyclic attractor will still be bisimulated by 
a cyclic attractor since, by definition of coding, a transition between two different integer 
states is always simulated by a 
transition with two different Boolean profiles.
Figure~\ref{fig:equilibria-different-nature} shows an example where the self-loop of stable state 
$2$ is simulated by a cyclic attractor over the three Boolean profiles coding for it (right-hand 
side 
network). Indeed, for any integer level, the synchronous dynamics allows reaching any of its 
codings from any other one. This case is however not encountered for stable state $3$ coded by a 
single Boolean 
profile. The occurrence of such situations also depends on the concrete Boolean function, as 
shown 
by the middle network that 
preserves the stability. Even though in the former case the stable state is represented by a 
cyclic 
attractor, it is worth noticing that the original level 2 can be recovered from the states 
encompassed by the attractor since 
$\{\psi(101),\psi(011),\psi(110)\} =\{2\}$. 

Maintaining the stability matters for the analysis performed on the Boolean dynamics. In 
particular, the symbolic computation of stable states will fail to find state $2$ as equilibrium 
from the Boolean network since this state is represented by a cyclic attractor. Therefore, such
cases should be ruled out to ensure a matching analysis of the dynamics on the two networks.
To preserve the stability of equilibria, we basically have to prevent reaching a code of an integer level $l$ from another
code also encoding for $l$.
This depends on the mode and on the Boolean function (\textit{cf.} Figure~\ref{fig:equilibria-different-nature}). The expected outcome can be expressed as follows for a mode $M$:
\begin{equation}
\forall w \in \dom \psi, \forall m \in M: w_m \neq f_m(w) \implies \psi(w) \neq \psi(f_m(w) \cup w_{-m}).
\label{eq:stability-preserving}
\end{equation}

We examine two effective conditions for satisfying~(\ref{eq:stability-preserving}) that are 
independent of the specification of the Boolean network. A simple one working whatever the mode and the 
Boolean function is to remove the self-loops, and thus establish bisimulation between reflexive 
reductions of the state graphs of both networks, instead of operating on the original state graphs. The equilibrium stability then remains 
preserved since no circuits can simulate a self-loop and the important features of the reachability 
are not altered.
Another more explicit condition, based on the code and the sizes of modalities, forbids the access by a transition to another Boolean profile 
coding for the same integer.
\begin{proposition}
	Let $B= \tuple{f,\sprt{Y},M}$ be a Boolean network bisimilar to a Multi-valued network 
	$N=\tuple{g,Y, \mathbf{1}_Y}$, with $M$ a local-to-support mode. If the following holds: 
	\begin{multline*}
	\forall y_i \in Y, \forall s_{y_i} \in \Nset_{y_i}, \forall w,w' \in \psi^{-1}(s_{y_i}): \\
	 w \neq w' \implies d(w,w') > \max\{ |m| \mid m \in M \},
	\end{multline*}
	then the equilibrium stability~(\ref{eq:stability-preserving}) is preserved. 
	\label{prop:no-reach-code-from-code}
\end{proposition}
 Consequently, if $\psi$ is a bijection, the stability of equilibria is preserved since every integer level is coded by a 
 single Boolean profile. On the other hand, the asynchronous mode preserves the stability under the Summing code, since 
 distances between two Boolean profiles coding for the same integer are at least $2$.

\subsection{Preservation of regulation}
\label{sec:regulation-preserving}
A Boolean network bisimulating a Multi-valued network is \emph{regulatory-preserving} if it is possible to unambiguously recover the signed interaction graph of the Multi-valued network (\textsc{migs}), $\tuple{Y,\blackarc,\sigma}$, from the signed Boolean interaction graph of the bisimilar Boolean network (\textsc{bigs}), $\tuple{\sprt{Y},\arc,\sigma_\Bset}$.
 Retrieving \textsc{migs} from \textsc{bigs} is divided in two steps: retrieving the interaction graph and finding the signs.

\paragraph{Interaction graph retrieval} The structure of \textsc{migs} is retrieved from the quotient graph of \textsc{bigs} defined on the support of the integer variables, called the \emph{support interaction graph} (\textsc{sig}) $\tuple{\{\sprt{y_i}\}_{y_i \in Y}, \blackarc}$, where an interaction between two Boolean support variables induces an interaction between the supports they belong: 

\begin{equation}
\sprt{y_i} \blackarc \sprt{y_i} \eqdef \exists \sprt{y_i}_k \in \sprt{y_i}, \exists \sprt{y_j}_{r} \in \sprt{y_j}: \sprt{y_i}_{k} \arc \sprt{y_j}_{r}.
\label{eq:support-interaction-graph-quotient}
\end{equation}
As a consequence, the topological structure of \textsc{migs} is the same as that of \textsc{sig} by merely replacing the supports by the integer variables they support (Proposition~\ref{prop:MIGS = SIG}). In fact, \textsc{sig} essentially provides an intermediary representation used for recovering the interactions of \textsc{migs} and their signs.
\begin{proposition} Let $N$ be a Multi-valued network and $B$ a Boolean network. If $ N$ is bisimilar to $B$ then $\operatorname{\textsc{migs}}(N)$ is isomorphic to $\operatorname{\textsc{sig}}(B)$.
	\label{prop:MIGS = SIG}
\end{proposition}

\paragraph{Sign retrieval}
The sign of an interaction is determined by \textsc{bigs} once the conversion is achieved (see 
Figures~\ref{fig:boolean-network-ex-sc}, \ref{fig:boolean-network-ex-gc}). Therefore the issue is to 
deduce from the signs of the interactions between the Boolean variables the signs of the corresponding interactions in \textsc{migs}. 
The recovery procedure is based on a set of reference Boolean variables, considered as 
\emph{markers of sign}, covering all the supports such that the signs of the interactions between 
these variables are the same as the signs of the interactions between the integer variables they 
support. 
Hence the set of markers $\mathcal{M}_{\sprt{Y}}$ is a subset of Boolean variables of 
$\sprt{Y}$ defined by:
\begin{definition}[Markers of sign]
	Let $\tuple{g,Y}$ a Multi-valued network bisimulating
	a Boolean network $\tuple{f,\sprt{Y}}$ with $\tuple{Y,\blackarc,\sigma}$ and $\tuple{\sprt{Y}, \arc, \sigma_\Bset}$ as their respective signed interaction graphs.
	$\mathcal{M}_{\sprt{Y}} \subseteq \sprt{Y}$ is a set of \emph{markers of sign} if and only if:
	\begin{enumerate}
		\item The sign $\sigma$ of an interaction between any two Boolean variables in $\mathcal{M}_{\sprt{Y}} \subseteq \sprt{Y}$ is the same as the sign of the interaction between the integer variables that they support:
		$$\forall \sprt{y_i}_k, \sprt{y_j}_r \in \mathcal{M}_{\sprt{Y}} : \sprt{y_i}_k \sarc{\sigma} \sprt{y_j}_r \iff y_i \sblackarc{\sigma} y_j. $$ 
		\label{item:markers-def-1}
		\item All integer variables have markers:
		$$\forall y_i\in Y: \mathcal{M}_{\sprt{Y}} \cap \sprt{y_i} \neq \emptyset.$$ 
		\label{item:markers-def-2}
	\end{enumerate}
	
	\label{def:markers}
\end{definition}

To operationally identify the markers from a code, we define a code-based marker condition (\ref{eq:markers-min-condition}) directly linking the markers to the code for the asynchronous mode. This condition asserts the monotony of the coding for markers with respect to the integer and Boolean orders by stipulating that an integer coded by a Boolean profile is less than another coded by this Boolean profile where a marker value is substituted by $1$
 (Lemma~\ref{lem:markers-specification-from-code}).

\begin{lemma}
	\label{lem:markers-specification-from-code}
	Let $N=\tuple{g,Y}$ be a Multi-valued network bisimulating
	an asynchronous Boolean network $B=\tuple{f,\sprt{Y},\mathbf{1}_{\sprt{Y}}}$, and $ \mathcal{M}_{\sprt{Y}} \subseteq \sprt{Y}$ be a set of Boolean variables complying to (\ref{def:markers}.\ref{item:markers-def-2}).  If: 
	\begin{equation}
	\forall \sprt{y_i}_k \in \mathcal{M}_{\sprt{Y}}, \forall w\in \dom{\psi}: \psi(w) \leq \psi(w_{[\sprt{y_i}_k \mapsto 1]})
	\label{eq:markers-min-condition}
	\end{equation}
	
	\noindent
	\medskip
	then $\mathcal{M}_{\sprt{Y}}$ fulfils Definition~(\ref{def:markers}.\ref{item:markers-def-1}) 
	and $\mathcal{M}_{\sprt{Y}}$ is a set of markers.
\end{lemma}
 Therefore, the goal is to determine for each integer variable the set of markers by checking~(\ref{eq:markers-min-condition}) for a given coding.
For the Summing code all the Boolean variables are markers, and for the Gray code the variables 
storing the most significant bit indexed by $1$ ($\sprt{y_i}_1$) are the markers. 

\pagebreak
\begin{theorem}
	Let $N=\tuple{g,Y}$ be a Multi-valued network in bisimulation with an asynchronous Boolean network $B=\tuple{f,\sprt{Y},\mathbf{1}_{\sprt{Y}}}$. The sets of markers $\mathcal{M}_{\sprt{Y}}$ are respectively for the codes:
	\begin{itemize}
		\item Summing code: $\mathcal{M}_{\sprt{Y}} = \sprt{Y}$;
		\item Van Ham code: $\mathcal{M}_{\sprt{Y}} = \sprt{Y}$;
		\item Gray code: $\mathcal{M}_{\sprt{Y}}= \{\sprt{y_i}_1 \mid y_i \in Y \}$.
	\end{itemize}	
	\label{thm:markers-characterization}
\end{theorem}

\section{Inference of Boolean formulas}
\label{sec:algo-conversion}

 An analytical definition of the Boolean network function is given by (\ref{eq:main-fg}). Although 
 the function $\psi^{-1}\circ g \circ \psi$ is closed on Boolean states when $\psi$ is bijective 
 characterizing a Boolean network, the Boolean formulas are not explicitly defined. 
 The lack of Boolean formulas makes the analysis harder in practice, notably by 
 preventing the characterization of the interaction graph directly from formula specifications. 
 Moreover, the analytical definition does not hold when  $\psi$ is not bijective, since $\psi^{-1}$ returns a 
 set of Boolean profiles. To circumvent this 
 limitation, the strategy is to infer the Boolean network bisimilar to a Multi-valued network 
 directly from the specification of the latter~(\ref{eq:Multi-valued-def}). As it is sufficient to find a  
 bisimilar Boolean network for each local Multi-valued evolution function $g_i$ 
 (Proposition~\ref{prop:local-to-global-bisimulation}), the algorithm will act on each function of 
 integer variables independently. 
 In this section we define a method inferring the formulas $f_{i,k}$ for each support variable $ 
 \sprt{y_i}_k$ of $y_i$ such that the reflexive reduction of the resulting Boolean network is 
 bisimilar to the reflexive reduction of the initial Multi-valued network, the code being a 
 parameter of this method. Due to the reflexive reductions, this method preserves the nature of the stable 
 states~(Section~\ref{sec:nature-equilibria-preserving}). 
 For simplicity, the inference is presented for the asynchronous mode, but it can be applied to any 
 local-to-support mode.  This point is discussed at the end of the section. 
 
 The definition of a formula $f_{i,k}$ for a support variable is divided in two stages: The \emph{conversion of the guard into a Boolean form}, and the \emph{derivation of the admissibility condition for guard validation}. The examples use the Summing code which is the most complex coding for the inference.

\paragraph{Boolean conversion of the guard} Basically, the guard of level $l'$ must also be 
satisfied in the Boolean network to simulate a transition shifting the current level $l$ to $l'$. 
The conversion of a Multi-valued guard to a \emph{Boolean guard} gathers the codes of the state profiles 
fulfilling the conditions of level $l'$, \ie $ \mathcal{C}_{ \star \shortblackarc 
y_i}^{l'}=\{ s_{ (\star \shortblackarc y_i)} \mid C^{l'}(s) \}$ where $(\star \shortblackarc y_i)$ 
is the set of regulators of $y_i$. As all these integer states satisfy the guard $C^{l'}$, their 
Boolean codes should also satisfy the Boolean guard $C^{l'}_{\Bset}$ defined as\footnote{The 
{minterm} of a state is a conjunction of the variables such that the unique interpretation satisfying 
it is the state itself, \eg $\minterm (x_1 =0, x_2=1)= \neg x_1 \land x_2$.}:
\begin{equation}
C^{l'}_{\Bset}=\bigvee_{s \in \mathcal{C}_{\star \shortblackarc y_i}^{l'}} \left (\bigwedge_{y_j 
\in (\star \shortblackarc y_i)} \bigvee_{w_{\sprt{y_j}} \in \psi^{-1}(s_{y_j})} \minterm 
(w_{\sprt{y_j}}) \right).
\label{eq:guard-booleanization}
 \end{equation}

For example, in the case of the Multi-valued network from Figure~\ref{fig:Multi-valued-network-ex}, the states fulfilling the conditions to reach level $2$ for $y$ are $(x=1, y=1)$ for the transition from level $1$ to $2$ for $y$, or $(x=0, y=3)$ for the transition from level $3$ to $2$. The code for $x$ is $\psi_x^{-1}(0)=\{(0)\}$, $\psi_x^{-1}(1)=\{(1)\}$ and the codes for $y$ are respectively for $1$ and $3$: $\psi_y^{-1}(1)=\{(0, 0, 1), (0, 1, 0), (1, 0, 0)\}$ and $\psi_y^{-1}(3)=\{(1, 1, 1)\}$. 
Hence, the Boolean guard of level $2$ for $y$ is: 

\begin{multline*}
C^{2}_{\Bset}= \\
\overbracket{
\left(\underbrace{\sprt{x}}_{\minterm_x (1)} \land \left(
 \underbrace{\left(\land \sprt{y}_1 \land \neg \sprt{y}_2 \land \neg \sprt{y}_3 \right)}_{\minterm_y (1, 0, 0)}
 \lor
 \underbrace{\left(\neg \sprt{y}_1 \land \sprt{y}_2 \land \neg \sprt{y}_3 \right)}_{\minterm_y (0, 1, 0)} \lor
 \underbrace{ \left(\neg \sprt{y}_1 \land \neg \sprt{y}_2 \land \sprt{y}_3 \right)}_{\minterm_y (0, 0, 1)}
 \right) \right)}^{(x=1, y=1)} 
\\ \lor 
\overbracket{\left(\underbrace{\neg \sprt{x}}_{\minterm_x (0)} \land \underbrace{\left( \sprt{y}_1 \land \sprt{y}_2 \land \sprt{y}_3 \right)}_{\minterm_y (1, 1, 1)}\right)
}^{(x=0, y=3)}
\end{multline*}

\paragraph{Guard admissibility condition} The generation of the Boolean guard is however insufficient for obtaining the final formula because some support variables shift to $0$ during the transition even though the guard is satisfied, meaning that a direct evaluation of the Boolean guard would shift them to $1$. For example, in Figure~\ref{fig:Multi-valued-network-ex}, shifting from $3$ to $2$ for $y$ is bisimilar to $(1, 1, 1) \xevo{\sprt{y}_2} (1, 0, 1)$. In this case, we need to shift the state of $\sprt{y}_2$ to $0$ although the guard is satisfied with $s_x=0$. We thus need to characterize the situations in which the transition necessarily shifts the value of a support variable to $1$. This restricts the set of admissible encodings triggering the guard, outside of which the transition always shifts the support variable state to $0$. 
 
\medskip
Let $s \xevo{y_i} s'$ be an integer unitary stepwise transition with $s_{y_i}=l$ and $s'(y_i)=l'$ such that $|l -l'| = 1$, and $\sprt{y}_{i}$ be the support of $y_i$ ($\sprt{y_i}_{k} \in \sprt{y_i}$), we denote by $w \evo w'$ the asynchronous Boolean transition bisimilar to $s \xevo{y_i} s'$. Two cases where $\sprt{y}_{i, k}=1$ should be considered depending on the encoding of the levels: $\sprt{y}_{i, k}$ is shifted from $0$ to $1$ during the transition (\ie $w_{\sprt{y}_{i, k}}=0$ and $w'(\sprt{y}_{i, k})=1$), or $\sprt{y}_{i, k}$ remains as $1$ (\ie $w_{\sprt{y}_{i, k}}=1$ and $w'(\sprt{y}_{i, k})=1$).

In both cases, we characterize for each Boolean variable the set of codes corresponding to the initial level $l$ such that $\sprt{y_i}_{k}$ is either shifted to or remains at $1$.																																																															 The initial level $l$ is determined from the target level $l'$ by considering that it is either $l'-1, l'$ or $l'+1$ by definition of an unitary stepwise transition.

\medskip
We define the set of codes for the initial level such that $\sprt{y_i}_{k}$ is shifted from $0$ to $1$ during the transition ($\psi$ is implicitly restricted to $\psi:\Bset_{\sprt{y_i}} \to \Nset_{y_i}$):
\begin{equation*}
\begin{multlined}
\Psi_{0\to 1}(l',\sprt{y_i}_{k}) = 
	\{ w_{\sprt{y_i}} \in \dom \psi \mid \exists \; \max(0, l'-1)\leq l \leq \min(l'+1, L): \\
	\psi(w_{\sprt{y_i}})=l \land w_{\sprt{y_i}_k}=0 \land 
	\psi({w_{\sprt{y_i}}}_{[\sprt{y_i}_{k} \mapsto 1]})=l' \}.
\end{multlined}
\end{equation*}
 Similarly, we define the set of codes for which a shift from $1$ to $0$ occurs:
\begin{equation*}
\begin{multlined}
\Psi_{1\to 0}(l',\sprt{y_i}_{k}) = 
\{ w_{\sprt{y_i}} \in \dom \psi\mid \exists \; \max(0, l'-1)\leq l \leq \min(l'+1, L): \\
\psi(w_{\sprt{y_i}})=l \land w_{\sprt{y_i}_k}=1 \land 
\psi({w_{\sprt{y_i}}}_{[\sprt{y_i}_{k} \mapsto 0]})=l' \}.
\end{multlined}
\end{equation*}
Finally, we define the set of codes where $\sprt{y_i}_{k}$ is $1$ in both $l$ and $l'$:
\begin{equation*}
\begin{multlined}
\Psi_{1\to 1}(l',\sprt{y_i}_{k}) = \{ w_{\sprt{y_i}} \in \dom \psi \mid \exists \; \max(0, 
l'-1)\leq l \leq \min(l'+1, L), \\ \exists w'_{\sprt{y_i}_k} \in \psi^{-1}(l') : 
\psi(w_{\sprt{y_i}})=l \land w_{\sprt{y_i}_k}=1 \land w'_{\sprt{y_i}_k}=1 \}.
\end{multlined}
\end{equation*}
The set of Boolean states coding for level $l$, always reaching state $1$ and never a state $0$ for $\sprt{y_i}_{k}$ in a transition to a code 
of level $l'$, is defined as:
\begin{equation*}
\Psi_{\star \to 1}(l',\sprt{y_i}_{k}) = \Psi_{0\to 1}(l',\sprt{y_i}_{k}) \; \cup \left(\Psi_{1\to 
1}(l',\sprt{y_i}_{k}) 
 \setminus \Psi_{1\to 0}(l',\sprt{y_i}_{k}) \right).
\end{equation*}
Note that the set difference in the previous equation is not
necessarily empty.  Indeed, there may exist a pair of states
$w_{\sprt{y_i}}$ and $w'_{\sprt{y_i}}$, with
$\psi(w_{\sprt{y_i}}) = l$ and $\psi(w'_{\sprt{y_i}}) = l'$, such that
$w_{\sprt{y_i}_k} = w'_{\sprt{y_i}_k} = 1$, but for which
$\psi({w_{\sprt{y_i}}}_{[\sprt{y_i}_{k} \mapsto 0]})=l'$ also holds.
We need to exclude such states $w_{\sprt{y_i}}$ from
$\Psi_{\star \to 1}(l',\sprt{y_i}_{k})$, because they still allow
reaching a Boolean profile coding for $l'$ by setting $\sprt{y_i}_{k}$
to 0. The resulting transition is bisimilar to an integer transition, and thus must be kept.

The \emph{guard admissibility condition} $G$ of $C^{l'}_\Bset$ is thus defined as the disjunction 
of the minterms of the admissible codes:

\begin{equation}
G_{\Psi_{\star \to 1}(l',\sprt{y_i}_{k}) }= \bigvee_{ c \in\Psi_{\star \to 1}(l',\sprt{y_i}_{k})} \minterm(c).
\label{eq:guard-admissibility-condition}
\end{equation}

In our running example, consider the levels that potentially reach level $2$ in a unitary 
stepwise 
transition (levels 1, 2, and 3). The final simplified formulas of the code admissibility 
conditions 
$G_{\Psi_{\star \to 1}(2, \sprt{y}_k)}, 1 \leq k \leq 3$, for each support variable are detailed 
in 
Table~\ref{tab:admissible-code-level-2}.

From these conditions (Table~\ref{tab:admissible-code-level-2}), we deduce that the asynchronous 
transitions  from level $3$ coded by $(1,1,1)$ to level $2$ all set to $0$ one of the Boolean 
support variables.  Indeed, the update of $\sprt{y}_1$ to $0$ leads to $(0, 1, 1)$ and similarly 
for $\sprt{y}_2$, $(1,0,1)$ and $\sprt{y}_3$, $(1,1,0)$ that all represent a Summing code of 
level 
$2$. 
\begin{table}[ht]
	$$
	\begin{array}{l @{\;=\;} r}
	G_{\Psi_{\star \to 1}(2, \sprt{y}_1)} &	\left(y_1\land \neg y_2\right)\lor \left(y_2\land 
	\neg 
	y_3\right)\lor \left(\neg y_2\land y_3\right) \\
	G_{\Psi_{\star \to 1}(2, \sprt{y}_2)} &	\left(y_1\land \neg y_3\right)\lor \left(\neg 
	y_1\land 
	y_3\right)\lor \left(y_2\land \neg y_3\right) \\
	G_{\Psi_{\star \to 1}(2, \sprt{y}_3)} &	\left(y_1\land \neg y_2\right)\lor \left(\neg 
	y_1\land 
	y_2\right)\lor \left(\neg y_2\land y_3\right) \\
	\end{array}	
	$$		
	\caption{ Guard admissibility condition for level $2$ of the support variables of $y$.}
	\label{tab:admissible-code-level-2}
\end{table}

\paragraph{Boolean formula of a support variable} The final formula $f_{i,k}$ for a support variable $\sprt{y}_{i, k}$ can be understood as the Boolean version of the guards restricted to the codings admissible for their triggering, defined as:
\begin{equation}
f_{i,k} = \bigvee_{1 \leq l \leq L} \left(C_{\Bset}^{l} \land G_{\Psi_{\star \to 1}(l,\sprt{y_i}_{k})} \right).
\label{eq:final-formula-unitary-stepwise}
\end{equation}

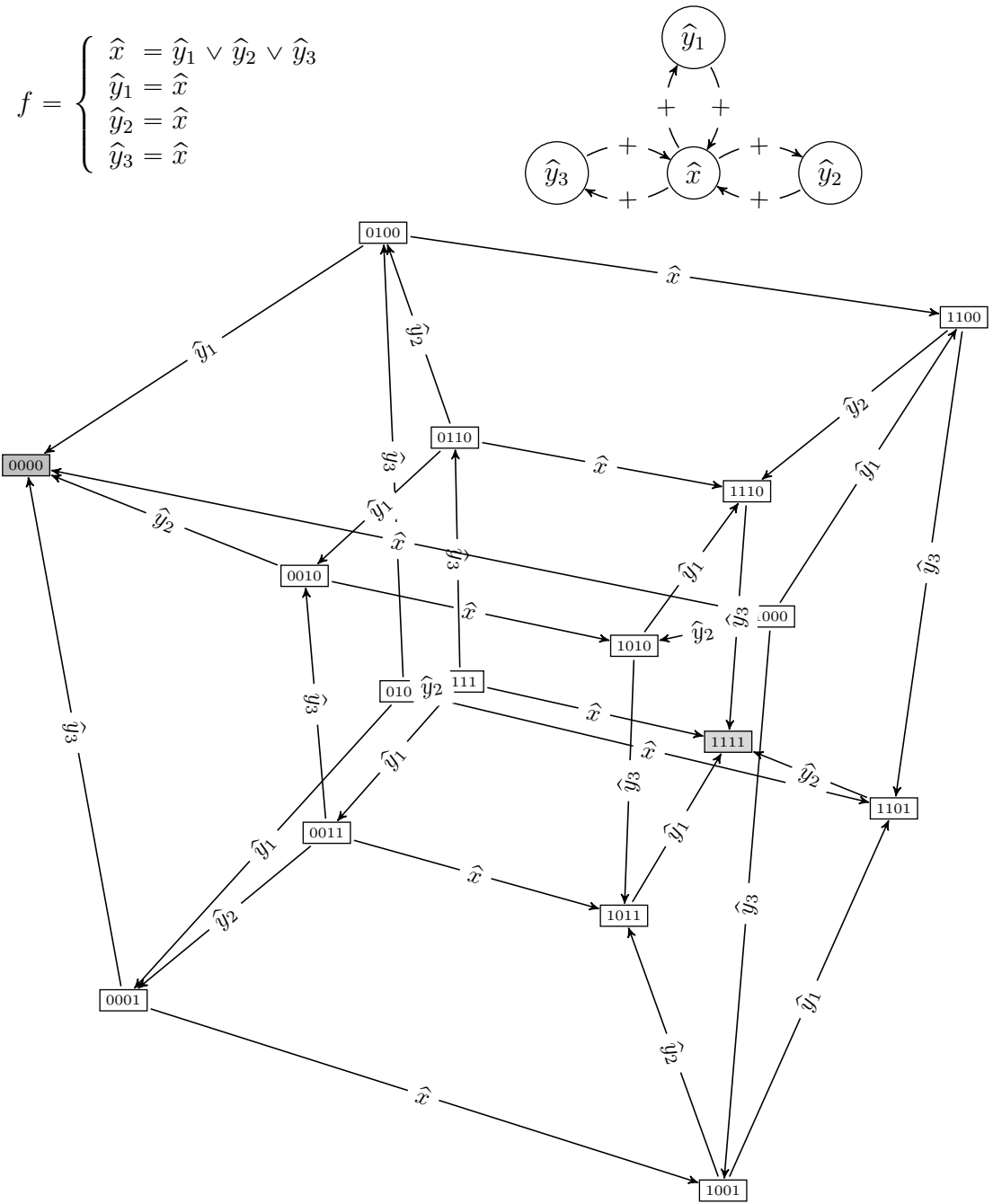
\begin{figure}[p]
	\begin{center}
		\begin{tabu} to \textwidth {X[l,$$,1] X[c,1]}
			f= \left\{
			\begin{array}{l @{\;=\;} l}
			\sprt{x} & \sprt{y}_1\lor \sprt{y}_2\lor \sprt{y}_3 \\
			\sprt{y}_1 & \sprt{x} \\
			\sprt{y}_2 & \sprt{x} \\
			\sprt{y}_3 & \sprt{x} 
			\end{array}
			\right.
			&
			\begin{tikzpicture}[baseline=5ex, x=2cm,y=2cm]
			\GraphInit[vstyle=Normal]
			\SetGraphUnit{1}
			\Vertex[L=$\sprt{x}$] {X}
			\NO[L=$\sprt{y}_1$](X){Y1}
			\EA[L=$\sprt{y}_2$](X){Y2}
			\WE[L=$\sprt{y}_3$](X){Y3}
			\SetUpEdge[style={->,bend left, thick, post,sloped}]
			
			\tikzstyle{LabelStyle}=[fill=white, sloped]
			\Edge[label=$+$](Y3)(X)
			\Edge[label=$+$](X)(Y2)
			\Edge[label=$+$](Y1)(X)
			\Edge[label=$+$](X)(Y1)
			\Edge[label=$+$](Y2)(X)
			\Edge[label=$+$](X)(Y3)
		\end{tikzpicture} 
	\end{tabu}
	
	\begin{tikzpicture}[baseline,xscale=2.5,yscale=2.4]
	\GraphInit[vstyle=Normal]
	\tikzset{VertexStyle/.append style={font=\tiny,draw,shape=rectangle,line width = 0.5,inner 
	sep = 2.5pt,outer sep = 0.5pt,minimum size = 0}}
	\tikzset{EdgeStyle/.style={post}}
	\tikzset{LabelStyle/.style={font=\footnotesize, sloped}} 
	\definecolor{colorof0000}{gray}{0.75} \tikzset{VertexStyle/.append style={fill=colorof0000}}
	\Vertex[x=0.15,y=4.63]{0000}
	\tikzset{VertexStyle/.append style={fill=white}}
	\definecolor{colorof1111}{gray}{0.85} \tikzset{VertexStyle/.append style={fill=colorof1111}}
	\Vertex[x=4.26,y=2.93]{1111}
	\tikzset{VertexStyle/.append style={fill=white}}
	\Vertex[x=0.72,y=1.34]{0001}
	\Vertex[x=1.78,y=3.95]{0010}
	\Vertex[x=1.91,y=2.37]{0011}
	\Vertex[x=2.24,y=6.06]{0100}
	\Vertex[x=2.36,y=3.24]{0101}
	\Vertex[x=2.66,y=4.8]{0110}
	\Vertex[x=2.69,y=3.3]{0111}
	\Vertex[x=4.51,y=3.7]{1000}
	\Vertex[x=4.23,y=0.17]{1001}
	\Vertex[x=3.71,y=3.52]{1010}
	\Vertex[x=3.65,y=1.86]{1011}
	\Vertex[x=5.64,y=5.54]{1100}
	\Vertex[x=5.23,y=2.52]{1101}
	\Vertex[x=4.37,y=4.47]{1110}
	\Edge[label={$\sprt{y}_3$}](0001)(0000)
	\Edge[label={$\sprt{x}$}](0001)(1001)
	\Edge[label={$\sprt{y}_2$}](0010)(0000)
	\Edge[label={$\sprt{x}$}](0010)(1010)
	\Edge[label={$\sprt{y}_2$}](0011)(0001)
	\Edge[label={$\sprt{y}_3$}](0011)(0010)
	\Edge[label={$\sprt{x}$}](0011)(1011)
	\Edge[label={$\sprt{y}_1$}](0100)(0000)
	\Edge[label={$\sprt{x}$}](0100)(1100)
	\Edge[label={$\sprt{y}_1$}](0101)(0001)
	\Edge[label={$\sprt{y}_3$}](0101)(0100)
	\Edge[label={$\sprt{x}$}](0101)(1101)
	\Edge[label={$\sprt{y}_1$}](0110)(0010)
	\Edge[label={$\sprt{y}_2$}](0110)(0100)
	\Edge[label={$\sprt{x}$}](0110)(1110)
	\Edge[label={$\sprt{y}_1$}](0111)(0011)
	\Edge[label={$\sprt{y}_2$}](0111)(0101)
	\Edge[label={$\sprt{y}_3$}](0111)(0110)
	\Edge[label={$\sprt{x}$}](0111)(1111)
	\Edge[label={$\sprt{x}$}](1000)(0000)
	\Edge[label={$\sprt{y}_3$}](1000)(1001)
	\Edge[label={$\sprt{y}_2$}](1000)(1010)
	\Edge[label={$\sprt{y}_1$}](1000)(1100)
	\Edge[label={$\sprt{y}_2$}](1001)(1011)
	\Edge[label={$\sprt{y}_1$}](1001)(1101)
	\Edge[label={$\sprt{y}_3$}](1010)(1011)
	\Edge[label={$\sprt{y}_1$}](1010)(1110)
	\Edge[label={$\sprt{y}_1$}](1011)(1111)
	\Edge[label={$\sprt{y}_3$}](1100)(1101)
	\Edge[label={$\sprt{y}_2$}](1100)(1110)
	\Edge[label={$\sprt{y}_2$}](1101)(1111)
	\Edge[label={$\sprt{y}_3$}](1110)(1111)
	\end{tikzpicture}
\end{center}
\caption{ A Boolean network bisimilar to the Multi-valued network in 
Figure~\ref{fig:Multi-valued-network-ex}, its interaction graph (right), and its asynchronous 
dynamics for the Summing code without the self-loops (below).}
\label{fig:boolean-network-ex-sc}
\end{figure}

The Boolean network gathers the formulas defined by (\ref{eq:final-formula-unitary-stepwise}) for 
each support variable. For the running example (Figure~\ref{fig:Multi-valued-network-ex}), the final 
Boolean network provides a clean description of the formulas once simplified for the Summing code 
(Figure~\ref{fig:boolean-network-ex-sc}) and the Gray code (Figure~\ref{fig:boolean-network-ex-gc}), 
that differ due to the codings. Theorem~\ref{thm:formula-inference-for-conversion} demonstrates 
the  correction of the conversion method.

\begin{figure}[h]
\begin{center}
	
	\begin{tabu} to 0.5\textwidth {X[c,1]}
		$
		\left\{
		\begin{array}{l}
		\sprt{x} = \sprt{y_1}\lor \sprt{y_2} \\
		\sprt{y_1} = \left(\sprt{x}\land \sprt{y_2}\right)\lor \left(\sprt{y_1}\land \neg
		\sprt{y_2}\right) \\
		\sprt{y_2} =\left(\sprt{x}\land \neg \sprt{y_1}\right)\lor \left(\neg \sprt{x}\land
		\sprt{y_1}\right) \\
		\end{array}
		\right.
		$
		\\
		\begin{tikzpicture}[baseline=2ex, x=2cm, y=3cm]
		\GraphInit[vstyle=Normal]
		\SetGraphUnit{1}
		\Vertex[L=$\sprt x$] {X}
		\SOWE[L=$\sprt{y}_1$](X){Y1}
		\SOEA[L=$\sprt y_2$](X){Y2}
		\SetUpEdge[style={->,bend left, thick, post,sloped}]
		\tikzstyle{LabelStyle}=[fill=white, sloped]
		
		\Edge[label=$\pm$](X)(Y2)
		\Edge[label=$+$](Y1)(X)

		\Edge[label=$+$](X)(Y1)
		\Edge[label=$+$](Y2)(X)
		\Edge[label=$\pm$](Y1)(Y2)
		\Edge[label=$\pm$](Y2)(Y1)
		\end{tikzpicture}
	\end{tabu}
	\hspace{1em}
	\begin{tikzpicture}[xscale=1.25,yscale=1.25, baseline={([yshift=-80pt] current bounding 
	box.north)}]]
	\GraphInit[vstyle=Normal]
	\tikzset{VertexStyle/.append style={font=\tiny,draw,shape=rectangle,line width = 0.5,inner 
	sep = 2.5pt,outer sep = 0.5pt,minimum size = 0}}
	\tikzset{EdgeStyle/.style={post}}
	\tikzset{LabelStyle/.style={font=\footnotesize, sloped}} 
	\definecolor{colorof000}{gray}{0.75} \tikzset{VertexStyle/.append style={fill=colorof000}}
	\Vertex[x=2.47,y=3.17]{000}
	\tikzset{VertexStyle/.append style={fill=white}}
	\definecolor{colorof110}{gray}{0.85} \tikzset{VertexStyle/.append style={fill=colorof110}}
	\Vertex[x=3.83,y=6.21]{110}
	\tikzset{VertexStyle/.append style={fill=white}}
	\Vertex[x=4.88,y=2.65]{001}
	\Vertex[x=2.39,y=5.83]{010}
	\Vertex[x=4.87,y=5.51]{011}
	\Vertex[x=3.84,y=3.93]{100}
	\Vertex[x=6.01,y=3.57]{101}
	\Vertex[x=6.07,y=6.]{111}
	\Edge[label={$\sprt{y}_2$}](001)(000)
	\Edge[label={$\sprt{x}$}](001)(101)
	\Edge[label={$\sprt{y}_2$}](010)(011)
	\Edge[label={$\sprt{x}$}](010)(110)
	\Edge[label={$\sprt{y}_1$}](011)(001)
	\Edge[label={$\sprt{x}$}](011)(111)
	\Edge[label={$\sprt{x}$}](100)(000)
	\Edge[label={$\sprt{y}_2$}](100)(101)
	\Edge[label={$\sprt{y}_1$}](101)(111)
	\Edge[label={$\sprt{y}_2$}](111)(110)
	\end{tikzpicture}		
\end{center}	
\caption{A Boolean network bisimilar to the Multi-valued network in 
Figure~\ref{fig:Multi-valued-network-ex}, its interaction graph (below), and its asynchronous 
dynamics for the Gray code without the self-loops (right).}
\label{fig:boolean-network-ex-gc}
\end{figure}

\paragraph{Bisimilar reflexive reduction}
Under the asynchronous mode, some support variables may maintain their value inducing self-loops that are not bisimilar to 
any transition in the integer dynamics. In the running example, shifting $y$ from $2$ to $3$ is bisimilar to 
$(1, 0, 1) \xevo{\sprt{y}_2} (1, 1, 1)$, which modifies the value of $\sprt{y}_2$. However, any of the 
Boolean variables may be updated in asynchronous dynamics leading to two self loops on $\sprt{y}_1$ 
and $\sprt{y}_3$ for maintaining the state of these variables at $1$. Obviously, these self-loops 
are not bisimilar to the integer state transition since the variation of the integer state from $2$ 
to $3$ is carried out by one transition only. No alternatives preventing these additional 
self-loops in the Boolean network are possible since any one of the Boolean variables may be updated, but
the state must not change for $\sprt{y}_1$ and $\sprt{y}_3$. This situation explains why our method operates on reflexive reductions of the 
networks, effectively discarding these extra self-loops. Also notice that this transformation can only be performed
if the integer level $0$ is coded by the $0$ Boolean profile, meaning that the behaviours of the Multi-valued and Boolean networks match when no guards are satisfied.
The reflexive reduction of a network $N$ is denoted ${N}^\neq$.

\pagebreak 

\begin{theorem}
	\label{thm:formula-inference-for-conversion}	
	Given a neighbourhood preserving Boolean coding $\psi$ such that $0$ is coded by the $0$ 
	Boolean profile,  the inference by (\ref{eq:final-formula-unitary-stepwise}) from a 
	Multi-valued unitary stepwise network $N=\tuple{g,Y, \mathbf{1}_Y}$  produces a  Boolean 
	network $B=\tuple{f, \sprt{Y}, \mathbf{1}_{\sprt{Y}}}$ such that the reflexive reductions of 
	both networks are bisimilar: $N^\neq \sim_\psi B^\neq$.
\end{theorem}

\paragraph{Extension to other modes} The method can be applied to any local-to-support mode but the 
admissible region may be reduced compared to the asynchronous mode. This reduction is caused by the 
decrease of the update capacity allowed by the mode. Hence, by Definition~\ref{def:bisimulation}, this implies selecting the 
appropriate codes for always reaching a code supporting an unitary stepwise transition for each 
update according to the mode components. For the running example, by using the 
parallel local-to-support mode with the Summing code, $M=\left\{\{x\},\{\sprt{y}_1,\sprt{y}_2, 
\sprt{y_3}\}\right\}$, all the Boolean variables of $\sprt{y}$ are updated jointly allowing to reach
a single code instead of reaching the different codes of the same integer by separate updates of 
variables. Thus, the coding is reduced to Van Ham coding.

Moreover, the trajectories starting from a state located outside the admissible region always end in the admissible region and no supplementary equilibra are thus created (Proposition~\ref{prop:reach-admissible-region}). This property generally holds for any coding that partially covers the Boolean state space.

In conclusion, the domain of $\psi$ may thus be reduced for a surjective decoding function such as the Summing code without altering the asymptotic dynamics, but remains unchanged for a bijective decoding.

\begin{proposition} Let $f$ be an evolution function defined according to 
(\ref{eq:final-formula-unitary-stepwise}) from a Multi-valued unitary stepwise network 
$N=\tuple{g,Y, \mathbf{1}_Y}$.  Let $B=\tuple{f,\sprt{Y},M}$ be the corresponding Boolean network with $M$ a local-to-support mode, and $\psi$ a decoding 
function such that $N^\neq \sim_\psi B^\neq$.
	 All the states of the Boolean space eventually reach a state in the admissible region:
	\begin{equation*}
	\forall w \in \Bset_{\sprt{Y}}, \exists w' \in \dom \psi: w \evo_f^\star w'. 
	\end{equation*}
	\label{prop:reach-admissible-region}
\end{proposition}

\paragraph{Complexity of the algorithm} Assume that the Multi-valued network has $n$ variables 
reaching at most level $L$, the upper bound on the number of regulators for a variable is $r$, the maximal 
number of support variables is $m$, and the maximal bound of code variants is $c$. Then the 
complexity of the Boolean guard is in $\mathcal{O}(n L^r r c m)$ and the complexity of computing the guard 
admissibility condition is in $\mathcal{O}(n L c m)$. Thus the complexity of the 
algorithm is dominated by the complexity of computing the Boolean guard.
 Accordingly, the computation time mainly depends on the maximal level and the number of regulators. The 
 algorithm is efficient in practice since the maximal level and the number of regulators often 
 remain tractable on real biological network models.

\section{Conclusion}
\label{sec:conclusion}
Bisimulation is at the core of the Multi-valued-to-Boolean network conversion process with the coding as a parameter. Although bisimulation preserves reachability, more features are expected to be preserved in order to directly perform analysis on the Boolean network. These features pertain to the nature of the equilibria and to the regulation of the variables. We propose a method whose outcome is a Boolean network explicitly defined by its Boolean formulas, with equilibria of the same nature as in the bisimilar Multi-valued network. The interaction graph of the Multi-valued network can be recovered from this Boolean network for any coding.
In particular, we show that the Gray code providing the shortest Boolean representation has the same properties as the Summing code, usually considered standard for this conversion.

Such automatic conversion sketches a pipeline where the Multi-valued network becomes an input
specification for modelling only, while the bulk of the analysis is performed on the Boolean
network. Such pipeline suggests that the Boolean framework is central and sufficient for biological 
network modelling, thus calling to focus theoretical efforts on this framework since the results 
will benefit to both categories of discrete models via this pipeline. 

 
 A perspective research direction would concern the study of bisimulation between Boolean networks. As 
 bisimulation formally represents a form of behavioural equivalence, we could investigate the global 
 properties of families of bisimilar Boolean networks in order to discover general rules governing 
 their behaviour. 

\vfill \clearpage 
 
\section*{Appendix}

\begin{proof}[Proposition~\ref{prop:localtoglobal}]
 The global Boolean transition relation is the union of the local transition relations
 that are bisimilar to the local Multi-valued relations. As the union of bisimilar relations is
 bisimilar to the union relation, we deduce that the global Boolean relation is bisimilar to the 
 global Multi-valued relation. 
\end{proof}

\begin{proof}[Theorem~\ref{thm:bisimulation-psi.f=g.psi-equivalence-local-to-support}]
	We prove that  Property~(\ref{eq:main-fg}) is met if and only if  $N\sim_\psi B$. We first 
	prove the implication and next the reciprocal. Before, we prove the following property 
	for any mode component $m \in M$ used in the proofs:
	\begin{equation}
		-m = (\sprt{y_i}\setminus m \cup -\sprt{y_i})
		\tag{T1}
		\label{eq:t1}
	\end{equation}
	\proof ~
	
	  \begin{tabu} to \textwidth {X[r,0.5,$] @{\;=\;} X[l,1.5,$]  X[l,4] }		
		-m &  \sprt{Y}\setminus m  & by definition of $-m$;\\
		& (\sprt{y_i} \cup -\sprt{y_i})\setminus m  & as $-\sprt{y_i}=\sprt{Y} \setminus 
		\sprt{y_i}$ by 
		definition; \\
		& \sprt{y_i}\setminus m \cup -\sprt{y_i}\setminus m \\
		& (\sprt{y_i}\setminus m \cup -\sprt{y_i}) &   since $m \subseteq \sprt{y_i}$ by 
		definition of the local-to-support mode, meaning that
		$ -\sprt{y_i}\setminus m = -\sprt{y_i}$. \\
		\end{tabu}  \qed
		
	\medskip	
 $(\implies)$
	Assume that Property~(\ref{eq:main-fg}) is met for the local-to-support mode $M$, \ie $\forall m
	\in M: m \subseteq \sprt{y_i} \land \psi(f_{m}(w) \cup w_{\sprt{y_i} \setminus m}) = g_i \circ 
	\psi(w)$.
	
	\begin{itemize}
		\item \textit{$N$ simulates $B$.} 
		Let $w \xevo{m}_f w', m \in M$, be a transition in the model of $B$ such that $w, w' \in 
		\dom \psi$.
		We define the transition $\psi(w) \evo \psi(w')$ by application of $\psi$ on $w$ and $w'$. We 
		have:
			
		\noindent
		\begin{tabu}{ X[l,$,0.75] @{$\;=\;$} X[l,$,4] X[l,4] }
			\psi(w') 
			& \psi( f_m(w) \cup w_{-m}) 
			& by definition of a transition (Section~\ref{sec:network-theory});\\
			& \psi( f_m(w) \cup w_{\sprt{y_i}\setminus m \cup -\sprt{y_i}})
			&  by~(\ref{eq:t1});\\
			& \psi( f_m(w) \cup w_{\sprt{y_i}\setminus m} \cup w_{-\sprt{y_i}}) 
			& from~(\ref{eq:support}) \\		
			& \psi( f_m (w) \cup w_{\sprt{y_i}\setminus m})  \cup \psi(w_{-\sprt{y_i}}) 
			& from (\ref{eq:support}) and (\ref{eq:support-fun});
			\\
			& g_i\circ \psi(w) \cup \psi(w_{-\sprt{y_i}})
			& from~(\ref{eq:main-fg}), true by hypothesis.
		\end{tabu}
		 
		 	\medskip
		 \noindent
		 Set $s= \psi(w)$ and $s'= \psi(w')$.
		 Then $s_{-y_i} = \psi(w_{-\sprt{y_i}})$, because $w_{-\sprt{y_i}}$ is the Boolean encoding 
		 of 
		 the 
		 rest of the state $s_{-y_i}$. We finally have: 
		 $$ \psi(w) \evo \psi(w') = s \evo g_i(s) \cup s_{-y_i} = s \xevo{y_i}_{g_i} s', $$
		 which defines a transition of $\evo_{g_i}$ with the asynchronous mode 
		 $\textbf{1}_{y_i}$.
		 
	   \item \textit{$B$ simulates $N$.}
		Let $s \xevo{y_i}_{g} s'$ be a transition in the model of $N$.
		As $ \psi:\Bset_{\sprt{Y}} \to \Nset_Y$ is surjective, there exist two Boolean states $w, 
		w'\in  \Bset_{\sprt{Y}}$ such that: $ \psi(w)=s$ and $ \psi(w')=s'$. We prove that we can select  $w'$ in the preimage of~$s'$ so that a transition $w \xevo{m}_f w'$ exists in the model of the Boolean network $B$. 
	
		\medskip
		\noindent
		Firstly, let us characterize $s'$ based on $w$.
		
		\noindent
		\begin{tabu} to \textwidth{ X[r,$,0.5] @{$\;=\;$} X[l,3,$] @{ } X[l,4] }
			s' & g_i(s) \cup s_{-y_i} 
			   & by definition of transition (Section~\ref{sec:network-theory});\\
			
		    	& g_i \circ \psi(w) \cup s_{-y_i} & as $ \psi(w)=s$ by hypothesis; \\
		
			     & \psi(f_m(w) \cup w_{\sprt{y_i} \setminus m}) \cup s_{-y_i} 
			     & from (\ref{eq:main-fg}), true d by hypothesis;\\
			
			&\psi(f_m(w) \cup w_{\sprt{y_i} \setminus m})  \cup \psi(w_{\sprt{-y_i}}) 
			& from (\ref{eq:support-fun}) and $\psi(w)=s$;\\
			
			&\psi(f_m(w) \cup w_{\sprt{y_i} \setminus m})  \cup \psi(w_{-\sprt{y_i}}) 
			& by definition of the support~(\ref{eq:support});\\
			
			& \psi(f_m(w) \cup w_{\sprt{y_i} \setminus m})  \cup w_{-\sprt{y_i}})
			& from  (\ref{eq:support-fun});\\
			
			& \psi(f_m(w) \cup w_{\sprt{y_i} \setminus m \cup -\sprt{y_i}}
			& by definition of the support~(\ref{eq:support});\\
			
			&  \psi(f_m(w) \cup w_{-m})  
			& by~(\ref{eq:t1}).
		\end{tabu}
		\medskip
		
		\noindent
		Thus, we conclude that $\psi(w') = s'$  implies that $w' =f_m(w) \cup w_{\sprt{y_i} 
		\setminus m}$. Hence, by definition of a transition,  we have $w \xevo{m}_{f} w'$, meaning that $B$ simulates $N$. 
    \end{itemize}
   In conclusion, if Property~\ref{eq:main-fg} is verified then networks $N$ and $B$ are  bisimilar with respect to $\psi$.

\medskip
 $(\impliedby)$ Assume that $N \sim_\psi B$. Hence, for all 
	transitions $w \xevo{m}_{f} w'$ such that $w, w' \in \dom \psi$, there exist $s, 
	s'\in\Nset_Y$ such that $s \evo_{g_i} s'$ and $s= \psi(w), s'= \psi(w')$. 
	
	From the bisimulation, we deduce that: 
	
	\noindent
	\begin{tabu} to \textwidth{ X[r,$,0.5] @{$\;=\;$} X[l,$,3] @{ } X[l,4] }
		s' & \psi(w') & by hypothesis; \\
		& \psi(f_{m}\cup w_{-m}) & by definition of  $w \xevo{m}_f w'$; 
		\\
		&\psi(f_{m} \cup w_{\sprt{y_i}\setminus m \cup -\sprt{y_i}}) 
		& by~(\ref{eq:t1}); \\
		&\psi(f_{m} \cup w_{\sprt{y_i}\setminus m} \cup w_{-\sprt{y_i}}) 
		& from~(\ref{eq:support});\\
		&\psi(f_{m} \cup w_{\sprt{y_i}\setminus m}) \cup \psi(w_{-\sprt{y_i}})   & by 
		(\ref{eq:support}), (\ref{eq:support-fun});\\
		& \psi(f_{m} \cup w_{\sprt{y_i}\setminus m})  \cup s_{-y_i} & as $s= \psi(w)$ by hypothesis.
	\end{tabu} 

\medskip
     From the definition of a transition, we deduce the following:
     
	\noindent
	\begin{tabu} to \textwidth { X[r,$,0.5]@{$\;=\;$} X[l,3,$] @{ } X[l,4]}
	s'
		& g_i(s) \cup s_{-y_i} & as $s \evo_{g_i} s'$ by hypothesis; \\
		& g_i\circ \psi(w) \cup s_{-y_i} & as $s= \psi(w)$ by the bisimulation hypothesis.
	\end{tabu}
	
	\medskip 
	\noindent
	As $-y_i \cap y_i = \emptyset$ because $-y_i = Y \setminus y_i$, we can simplify the equation 
	by removing $s_{-y_i}$ in both part, leading to:
	$$ \psi(f_{m} \cup w_{\sprt{y_i}\setminus m})  = g_i\circ \psi(w) \text{ for all $w \in \dom 
	\psi$}, $$
 which defines Property~(\ref{eq:main-fg}).
\end{proof}

\begin{proof}[Lemma~\ref{lem:adm-mode-eq}]
 That admissibility for modes is reflexive and symmetric follows
 directly from Definition~\ref{def:adm-mode}. To show transitivity
 of admissibility for modes, consider three arbitrary modes $M_1, 
 M_2, M_3\subseteq 2^{\sprt{Y}}$, such that both $adm_{ \psi}(M_1, M_2)$
 and $adm_{ \psi}(M_2, M_3)$ (with respect to the functional
 bisimulation $B\sim_{ \psi} N$). We can show that clause (1) of
 Definition~\ref{def:adm-mode} is satisfied for modes $M_1$ and $M_3$
 in the following way:
 \[
 \setlength{\arraycolsep}{1mm}
 \begin{array}{rrll}
 & & adm_{ \psi}(M_1, M_2) \land adm_{ \psi}(M_2, M_3) &\\

 \implies & & \big(\forall m_2\in M_2, \exists m_1\in M_1 \, :\, adm_{ \psi}(m_1, m_2)\big)
 & \\
 & \land & \big(\forall m_3\in M_3, \exists m_2\in M_2 \, :\, adm_{ \psi}(m_2, m_3)\big)
 & \mbox{Definition~\ref{def:adm-mode} (1)}\\

 \implies & & \, \forall m_3\in M_3, \exists m_2\in M_2, \exists m_1\in M_1 \, : & \\
 && \;\;\, adm_{ \psi}(m_2, m_3) \land adm_{ \psi}(m_1, m_2) & \\

 \implies & & \, \forall m_3\in M_3, \exists m_1\in M_1\, :\, adm_{ \psi}(m_1, m_3), &
 \end{array}
 \]
 where the last transition is done by the symmetricity and
 transitivity of admissibility for modalities. Showing that clause
 (2) of Definition~\ref{def:adm-mode} is satisfied for $M_1$ and
 $M_3$ can be done symmetrically, which implies
 $adm_{ \psi}(M_1, M_3)$ and the transitivity of admissibility for
 modes.
\end{proof}

\begin{proof}[Theorem~\ref{thm:adm-bisimulation}]
 Consider the Boolean network $B=\tuple{f, \sprt{Y}, M_0}$ and the
 Multi-valued network $N=\tuple{g_i, Y, \textbf{1}_{y_i}}$, related by
 the bisimulation $B\sim_{ \psi} N$. Pick an $M_0$-admissible mode
 $M\subseteq 2^{\sprt{Y}}$ and consider the Boolean network
 $B'=\tuple{f, \sprt{Y}, M}$. We will show that $B'$ bisimulates $N$, 
 $B'\sim_{ \psi} N$, by directly checking clauses (1) and (2)
 of the definition of bisimulation
 (Definition~\ref{def:bisimulation}).

 \medskip\noindent
 {\it Clause (1) (forward simulation):} Take two Boolean states $w, 
 w'\in \dom \psi$ such that $w\xevo{m}_f w'$ for some $m\in M$.
 We can then write the following deduction:
 \[
 \begin{tabu} to \textwidth {X[r,1,$] X[l,7,$] X[l,3,$] }
 & w \xevo{m}_f w' & \\

 \implies & \exists m_0\in M_0, \, \exists w''\in\dom \psi &\\
 & \;\;w\xevo{m_0}_f w'' \land \psi(w') = \psi(w'') &
 \mbox{ $M$ is $M_0$-admissible} \\

 \implies & \exists m_0\in M_0, \, \exists w''\in\dom \psi &\\
 & \;\; \psi(w)\xevo{\mu(m_0)}_g \psi(w'') \land \psi(w') = \psi(w'') &
 \mbox{ $B\sim_{ \psi} N$} \\

 \implies & \exists m_0\in M_0, \, \exists w''\in\dom \psi &\\
 & \;\; \psi(w)\xevo{\mu(m_0)}_g \psi(w').
 \end{tabu}
 \]
 Remark that since $N$ is only allowed to update one variable, $y_i$, 
 $\mu(m_0)$ can only be equal to $\{y_i\}$.

 \medskip\noindent {\it Clause (2) (backward simulation):} Take any
 two integer states $s, s'\in \Nset_Y$ and an arbitrary Boolean
 state $w\in \Bset_X$. Since $N$ is only allowed to update $y_i$, 
 we can carry out the following deduction:
 $$
 \begin{tabu} to \textwidth {X[r,1,$] X[l,6,$] X[l,2,$]}
 & \psi(w) = s\land s\xevo{y_i}_g s' & 
 \\
 \implies & \exists w'\in \Bset_X, \, \exists m_0\in M_0 \,: \mu(m_0) = \{y_i\}\land \psi(w') = s'\land w\xevo{m_0}_f w' & B\sim_{ \psi} N \\

 \implies & \exists w'\in \Bset_X, \, \exists m_0\in M_0, 
 \exists w''\in \Bset_X, \exists m\in M \,: \mu(m_0) = \{y_i\}\land \psi(w') = s'\land w\xevo{m_0}_f w' 
 \land\, \psi(w'') = \psi(w')\land w\xevo{m}_f w'' & M \text{ is } M_0\text{-admissible} \\

 \implies & \exists w''\in \Bset_X, \, \exists m\in M \,:
 \psi(w'') = s'\land w\xevo{m}_f w''.
 \end{tabu}
$$

 \medskip \noindent
 The two previous paragraphs show that the clauses of the definition
 of bisimulation (Definition~\ref{def:bisimulation}) are satisfied
 for the Boolean network $B'$, running under mode $M$, and for the
 Multi-valued network $N$, meaning that $B'\sim_{ \psi} N$. The
 associated function mapping the modalities of $B'$ to those of $N$
 is the unique total function $2^{\sprt{Y}}\to \{y_i\}$ (i.e., the same
 as for the bisimulation $B\sim_{ \psi} N$).
\end{proof}
\begin{proof}[Proposition~\ref{prop:no-reach-code-from-code}]

By definition of a transition (Section~\ref{sec:network-theory}), we have:
$$\forall w,w' \in \Bset_{\sprt{Y}}, \forall m \in M: w \xevo{m} w' \implies d(w,w') \leq |m|,$$

\noindent
As by hypothesis,
$$ \forall y_i \in Y, \forall s_{y_i} \in \Nset_{\sprt{y_i}}, \forall v,v' \in \psi^{-1}(s_{y_i}): 
v \neq v' \implies \gamma < d(v,v'),$$
where $\gamma$ stands here for the greatest cardinality of $M$ components, 
$$\gamma= \max \{|m|, m \in M\},$$
 we deduce that: $d(w,w')\leq |m| \leq \gamma < d(v,v')$, meaning that the transition cannot be 
 achieved between codes of the same integer by hypothesis, thus leading to:
$$\forall w,w' \in \Bset_{\sprt{Y}}, \forall m \in M: w \neq w' \land w \xevo{m} w' \implies \psi(w) \neq \psi(w'),$$
As $w'=f_m(w) \cup w_{-m}$ by definition of a transition, this statement is equivalent to (\ref{eq:stability-preserving}), concluding that the equilibrium stability is preserved.
\end{proof}

\begin{proof}[Lemma~\ref{lem:markers-specification-from-code}]
	let $N$ be an asynchronous Multi-valued network bisimulating an asynchronous Boolean network $B$ with:
	
	\noindent
	$\operatorname{\textsc{migs}}(N)=\tuple{Y,\blackarc,\sigma}$, and $\operatorname{\textsc{bigs}}(B)=\tuple{\sprt{Y},\arc,\sigma_\Bset}$,
	as their respective signed interaction graphs; 
	let $ \mathcal{M}_{\sprt{Y}} \subseteq \sprt{Y}$ be a set of Boolean variables complying to~(\ref{eq:markers-min-condition}), we prove Statement~(\ref{def:markers}.\ref{item:markers-def-1}) by considering that Statement~(\ref{def:markers}.\ref{item:markers-def-2}) holds.

	First we demonstrate two properties (\ref{eq:proof-lem-markers-1}) and (\ref{eq:proof-lem-markers-2}) used in the proof:
	\begin{multline}
 \forall w,w' \in \dom \psi, 	\forall \sprt{y_i}_k \in \mathcal{M}_{\sprt{Y}}: \\
	w_{\sprt{y_i}_k} \leq w'_{\sprt{y_i}_k} \land w_{- \sprt{y_i}_k} = w'_{- \sprt{y_i}_k} \implies \psi(w) \leq \psi(w').
	\tag{L2.a}
	\label{eq:proof-lem-markers-1}
	\end{multline}
	
	\proof  Assume that:
		
		\noindent
		$\forall w,w' \in \dom \psi: w_{\sprt{y_i}_k} \leq w'_{\sprt{y_i}_k} \land w_{- \sprt{y_i}_k} = w'_{- \sprt{y_i}_k}$ for $\sprt{y_i}_k \in \mathcal{M}_{\sprt{Y}}$.
		
		\noindent
		Two cases occur: 
		\begin{enumerate}
			\item $w_{\sprt{y_i}_k} = w'_{\sprt{y_i}_k}$: in this case $w=w'$ leading to $\psi(w)=\psi(w')$ since $\psi$ is a function, thus satisfying $ \psi(w) \leq \psi(w')$.
			\item $w_{\sprt{y_i}_k} < w'_{\sprt{y_i}_k}$: as only two values are possible, $0$ or $1$, we 
			deduce that $ w'_{\sprt{y_i}_k}=1$. Hence, $w'$ can be defined as $w'= w_{[\sprt{y_i}_k 
			\mapsto 1]}$. As $\sprt{y_i}_k \in \mathcal{M}_{\sprt{Y}}$ by hypothesis, we conclude 
			from~(\ref{eq:markers-min-condition}) that $ \psi(w) \leq \psi(w_{[\sprt{y_i}_k \mapsto 
			1]})$. This inequality is equivalent to $\psi(w) \leq \psi(w')$. \qed
		\end{enumerate}

	\begin{multline}
	\tag{L2.b}
    \forall w,w' \in \dom \psi:
	w_{\sprt{y_i}_k} \leq w'_{\sprt{y_i}_k} \land w_{-\sprt{y_i}_k} = 
	w'_{-\sprt{y_i}_k} \implies \\
	\psi(w_{\sprt{y_i}}) \leq \psi(w'_{\sprt{y_i}}) \land \psi(w_{-\sprt{y_i}}) = \psi(w'_{-\sprt{y_i}}).
	\label{eq:proof-lem-markers-2} 
	\end{multline}
	\proof
		Assume that: $\forall w,w' \in \dom \psi: w_{\sprt{y_i}_k} \leq w'_{\sprt{y_i}_k} \land w_{-\sprt{y_i}_k} = w'_{-\sprt{y_i}_k}.$
		
		\noindent
		As $w_{-\sprt{y_i}} \subseteq w_{-\sprt{y_i}_k}$ and since $\sprt{y_i}_k \in \sprt{y_i}$, we have: 
		$w_{-\sprt{y_i}_k}= w'_{-\sprt{y_i}}\implies w_{-\sprt{y_i}}= w'_{-\sprt{y_i}},$ 
		
		\noindent
		thus implying that:
		$ \forall w,w' \in \dom \psi: w_{\sprt{y_i}_k} \leq w'_{\sprt{y_i}_k} \land w_{-\sprt{y_i}} = w'_{-\sprt{y_i}}.$
		
		\medskip
		\noindent
		Hence, from Property~(\ref{eq:proof-lem-markers-1}) applied to $w_{\sprt{y_i}}, w'_{\sprt{y_i}}$, we deduce that: 
		$$\forall w,w' \in \dom \psi: w_{\sprt{y_i}_k} \leq w'_{\sprt{y_i}_k} \land w_{-\sprt{y_i}} = w'_{-\sprt{y_i}} \implies \psi(w_{\sprt{y_i}}) \leq \psi(w'_{\sprt{y_i}}),$$
		Moreover, as $\psi$ is a function defined on supports, we have: 
		\begin{equation*}
		\forall w,w \in \dom \psi: w_{-\sprt{y_i}} = w'_{-\sprt{y_i}} \implies \psi( w_{-\sprt{y_i}}) = \psi(w'_{-\sprt{y_i}})
		\end{equation*}
		In conclusion, the following statement holds:
		$$\psi(w_{\sprt{y_i}}) \leq \psi(w'_{\sprt{y_i}}) \land \psi(w_{-\sprt{y_i}}) = 
		\psi(w'_{-\sprt{y_i}}).$$
		 \qed

	\noindent
	\paragraph{Now we prove that Statement~\ref{def:markers}.\ref{item:markers-def-1} is satisfied} The proof is given for positive interaction.

\medskip	
 $(\implies)$
	By definition~(\ref{eq:interaction}), a positive interaction, $\sprt{y_i}_k \sarc{+} \sprt{y_j}_r$ is defined as:
	$$ \forall w,w' \in \dom \psi: w_{\sprt{y_i}_k} \leq w'_{\sprt{y_i}_k} \land w_{-\sprt{y_i}_k} = w'_{-\sprt{y_i}_k} \implies f_{j,r}(w) \leq f_{j,r}(w').
	$$
	
	\noindent
	From (\ref{eq:proof-lem-markers-2}), we can rewrite this statement as:
	$$ \forall w,w' \in \dom \psi: \psi(w_{\sprt{y_i}}) \leq \psi(w'_{\sprt{y_i}}) \land \psi(w_{-\sprt{y_i}}) = \psi(w'_{-\sprt{y_i}}) \implies f_{j,r}(w) \leq f_{j,r}(w').$$
	
	\noindent
	Let $v = f_{j,r}(w) \cup w_{-\sprt{y_j}_r}$ and $v'= f_{j,r}(w') \cup w'_{-\sprt{y_j}_r}$, as $f_{j,r}(w) \leq f_{j,r}(w')$ by hypothesis, we conclude that: $\psi(v) \leq \psi(v')$ from (\ref{eq:proof-lem-markers-1}), thus leading to:
	\begin{multline*}
	\forall w,w' \in \dom \psi: \psi(w_{\sprt{y_i}}) \leq \psi(w'_{\sprt{y_i}}) \land \psi(w_{-\sprt{y_i}}) = \psi(w'_{-\sprt{y_i}}) \implies \\
	\psi( f_{j,r}(w) \cup w_{-\sprt{y_j}_r}) \leq \psi(f_{j,r}(w') \cup w'_{-\sprt{y_j}_r}).
	\label{eq:proof-lem-markers-key-statement}
	\end{multline*}
	
	As $N$ and $B$ are bisimilar, Property~(\ref{eq:main-fg}) holds. By application of this property we have: $\psi( f_{j,r}(w) \cup w_{-\sprt{y_j}_r})= g_j\circ \psi(w)$ and similarly for $w'$. Thus we deduce that: 
	\begin{multline*}
	\forall w,w' \in \dom \psi: \psi(w_{\sprt{y_i}}) \leq \psi(w'_{\sprt{y_i}}) \land \psi(w_{-\sprt{y_i}}) = \psi(w'_{-\sprt{y_i}}) \implies \\
	g_j\circ \psi(w) \leq g_j\circ \psi(w').
	\end{multline*}
	
	\noindent
	Finally, as $\codom \psi = \Nset_Y$ by definition, we can rewrite the previous statement as follows by setting, 
	$s=\psi(w),s'=\psi(w')$ :
	
	$$\forall s,s' \in \Nset_Y: s_i\leq s'_i \land s_{-i} = s'_{-i} \implies g_j(s) \leq g_j(s'),$$
	that defines the positive interaction on \textsc{migs}(N): $y_i \sblackarc{+} y_j$. 
	
	\medskip
 $(\impliedby)$
	Assume that an interaction $y_i \sblackarc{+} y_j$ exists and there exist two Boolean variables $\sprt{y_i}_k \in \mathcal{M}_{\sprt{Y}} \cap \sprt{y_i}$ and $\sprt{y_j}_r \in \mathcal{M}_{\sprt{Y}} \cap \sprt{y_j}$ with no positive interactions between these variables, \ie $\sprt{y_i}_k \sarc{\sigma} \sprt{y_j}_r \implies \sigma \neq +$.  We give a proof for the case $\sigma = -$; the proof for $\sigma = 0$ is similar.
	
	\medskip
	\noindent
	From definition of the interactions~(\ref{eq:interaction}), we deduce that:
	$$ \exists w,w' \in \dom \psi: w_{\sprt{y_i}_k} \leq w'_{\sprt{y_i}_k} \land w_{- \sprt{y_i}_k} = w'_{-\sprt{y_i}_k} \land 
	f_{j,r}(w) > f_{j,r}(w').$$
	From Property~(\ref{eq:proof-lem-markers-2}), we can rewrite the previous statement as:
	$$ \exists w,w' \in \dom \psi: \psi(w_{\sprt{y_i}}) \leq \psi(w'_{\sprt{y_i}}) \land \psi(w_{-\sprt{y_i}}) = \psi(w'_{-\sprt{y_i}}) \land f_{j,r}(w) > f_{j,r}(w').$$

	\noindent
	Let $v = f_{j,r}(w) \cup w_{-\sprt{y_j}_r}$ and $v'= f_{j,r}(w') \cup w'_{-\sprt{y_j}_r}$, as $f_{j,r}(w) > f_{j,r}(w')$ by hypothesis, we conclude that: $\psi(v) > \psi(v')$ from (\ref{eq:proof-lem-markers-1}), thus leading to:
	\begin{multline*}
	\forall w,w' \in \dom \psi: \psi(w_{\sprt{y_i}}) \leq \psi(w'_{\sprt{y_i}}) \land \psi(w_{-\sprt{y_i}}) = \psi(w'_{-\sprt{y_i}}) \implies \\
	\psi( f_{j,r}(w) \cup w_{-\sprt{y_j}_r}) > \psi(f_{j,r}(w') \cup w'_{-\sprt{y_j}_r}).
	\end{multline*}
	
	As $N$ and $B$ are bisimilar, Property~(\ref{eq:main-fg}) holds. By application of this property we have: $\psi( f_{j,r}(w) \cup w_{-\sprt{y_j}_r})= g_j\circ \psi(w)$ and similarly for $w'$. Thus we have:
	\begin{equation*}
	\exists w,w' \in \dom \psi: \psi(w_{\sprt{y_i}}) \leq \psi(w'_{\sprt{y_i}}) \land \psi(w_{-\sprt{y_i}}) = \psi(w'_{-\sprt{y_i}}) \land g_j\circ \psi(w) > g_j\circ \psi(w').
	\end{equation*}
	\noindent
	As $\codom \psi=\Nset_Y $ by definition, we can rewrite the previous statement as follows by setting, $s=\psi(w),s'=\psi(w')$ :
	
	$$\exists s,s' \in \Nset_Y: s_i \leq s'_i \land s_{-i} = s'_{-i} \land g_j(s) > g_j(s'),$$
	that contradicts the existence of a positive interaction $y_i \sblackarc{+} y_j$, which is false by hypothesis.
	
 \medskip
 The proof for negative interaction follows the same scheme. 
	Thus, we conclude that Statement~\ref{def:markers}.\ref{item:markers-def-1} is satisfied.
\end{proof}
\begin{proof}[Theorem~\ref{thm:markers-characterization}]
 We prove that~(\ref{eq:markers-min-condition}) holds for a set of Boolean variables belonging to a 
 support $\sprt{y_i}$.
 
 \medskip
	Let $\sprt{y_i}_k$ be a Boolean variable of this set, two cases occur: either $w_{\sprt{y_i}_k} 
	= 0$, or $w_{\sprt{y_i}_k}=1$. For the latter, $w$ is left untouched by substitution leading to 
	$\psi(w)=\psi(w_{[\sprt{y_i}_k \mapsto 1]})$ since $\psi$ is a function, thus 
	fulfilling~(\ref{eq:markers-min-condition}). Hence, we address the case when $w_{\sprt{y_i}_k} 
	= 0$ in the proofs.
	
	\paragraph{Summing code} The following property holds when $w_{\sprt{y_i}_k} = 0$:
	$$ \sum_{\sprt{y_i}_{j} \in \sprt{y_i} \setminus \sprt{y_i}_k} w_{\sprt{y_i}_j}= 
	 \sum_{\sprt{y_i}_j \in \sprt{y_i}} w_{\sprt{y_i}_j},$$ 
	 thus, we have:
	$$\psi(w_{[\sprt{y_i}_k \mapsto 1]}) = \sum_{\sprt{y_i}_{j} \in \sprt{y_i} \setminus 
	\sprt{y_i}_k} w_{\sprt{y_i}_j} +1 = \sum_{\sprt{y_i}_{j} \in \sprt{y_i}} w_{\sprt{y_i}_j} + 1= 
	\psi(w)+1.$$ 
	We conclude that: $\psi(w) < \psi(w_{[\sprt{y_i}_k \mapsto 1]}$.

	\paragraph{Van Ham code} Van Ham code is a sub-code of the Summing code, thus complying to its 
	results.
	
	\paragraph{Gray code} Let $\sprt{y_i}_1$ be a Boolean variable carrying the most significant digit, we separate ${\sprt{y_i}_1}$ from the other variables in the definition of $\psi$:
	
	$$
	\psi(w) = \sum_{k=1}^{|\sprt{y_i}|} 2^{|\sprt{y_i}| - k }.\bigoplus_{j=1}^{k} w_{\sprt{y_i}_{j}}= 
	2^{|\sprt{y_i}| -1 }.w_{\sprt{y_i}_{1}} + \sum_{k=2}^{|\sprt{y_i}|} 2^{|\sprt{y_i}| - k }.\bigoplus_{j=1}^{k} w_{\sprt{y_i}_{j}}. 
	$$

	Hence, when $w_{\sprt{y_i}_1} = 0$, we deduce that $\psi(w_{[\sprt{y_i}_1 \mapsto 1]}) = 
	2^{|\sprt{y_i}| -1 } +\psi(w)$, leading to $\psi(w) < \psi(w_{[\sprt{y_i}_1 \mapsto 1]})$.

\medskip	
Thus we conclude that $\sprt{Y}$ is the set of markers for the Summing and Van Ham code, while
$\{ \sprt{y_i}_1 \mid y_i \in Y \}$ are the markers for Gray code by application of 
Lemma~\ref{lem:markers-specification-from-code}.
\end{proof}	
\begin{proof}[Theorem~\ref{thm:formula-inference-for-conversion}]
 We first show that the computation of the Boolean function $f$ defined by~(\ref{eq:final-formula-unitary-stepwise}) is correct with respect to the integer function $g$ and the asynchronous mode (A).
 Next (B), we examine the satisfaction of Property~(\ref{eq:main-fg}).
 Finally we demonstrate the bisimulation of the reflexive reduction for both networks (C).
 
 \medskip \noindent
 \textit{A) The construction of $f$ is correct.}
 
	Let $s \xevo{y_i} s'$ be a Multi-valued transition, such that $s'_{y_i}= g_i(s)=l'$ and $s'_{y_j}=s_{y_j}$ for all $1 \leq j \leq n, j \neq i$, by definition of the asynchronous dynamics.  We have: $
	\max(l'-1,0) \leq s_{y_i} \leq \min(l'+1,L_i)$ since the evolution is unitary stepwise. There exist two Boolean states $w, w' \in \Bset_{\sprt{Y}}$ such that $\psi(w)=s$ as $\mathbf{codom}\;\psi = \Nset_Y$. We check that for all $\sprt{y_i}_k \in \sprt{y_i}$ if $f_{i,k}(w)=w'_{\sprt{y_i}_k}$ then $\psi(w') =s'$ and $w_{-\sprt{y_i}_k}= w'_{-\sprt{y_i}_k}$, thus proving the correction of $f_{i,k}$. The fact that $w_{-\sprt{y_i}_k}= w'_{-\sprt{y_i}_k}$ is a direct consequence an asynchronous transition updating one variable only.
	
	Two cases are considered qualifying whether $s'_{y_i} \neq 0$ or $s'_{y_i} = 0$.
	 For each, we examine whether the target state of the support variable $\sprt{y_i}_k$ is $0$ or $1$. 
	 Let us consider the following cases:	
\begin{enumerate}
\item $s'_{y_i} \neq 0$: 
 By definition of a Multi-valued network~(\ref{eq:Multi-valued-def}) $C_{l'}(s)$ is necessary satisfied as $l' = s'_{y_i} \neq 0$. Let $R(y_i)$ be the set of regulators of $y_i$, we have: $s_{R(y_i)} \in \mathcal{C}_{R(y_i),l'}$. Hence, we deduce that $w_{\sprt{y_i}}$ satisfies the Boolean version of the condition, $C^\Bset_{l'}$, by construction of the Boolean condition~(\ref{eq:guard-booleanization}).
Now we examine, the possible target states of the support variable $\sprt{y_i}_k$, $w'_{\sprt{y_i}_k}$:
\begin{itemize}
	\item $w'_{\sprt{y_i}_k}=1$: in this case $w_{\sprt{y_i}}$ belongs to $\Psi_{\star \to 1}(l',\sprt{y_i}_{k})$ by definition, meaning that $w$ admissible for the guard. Thus we have: 
	 $$f_{i,k}(w)=C_{l'}^{\Bset}(w) \land C_{\Psi_{\star \to 1}(l',\sprt{y_i}_{k})}(w)=1.$$
	\item $w'_{\sprt{y_i}_k}=0$: in this case $w_{\sprt{y_i}}$ does not belong to $\Psi_{\star \to 1}(l',\sprt{y_i}_{k})$ by definition meaning that $w$ is not admissible for the guard. Thus we have: 
	$$f_{i,k}(w)=C_{l'}^{\Bset}(w) \land C_{\Psi_{\star \to 1}(l',\sprt{y_i}_{k})}(w)=0.$$
\end{itemize}
In both cases, $f_{i,k}$ provides the expected result. 

\item $s'_{y_i}=0$: By definition of the Multi-valued dynamics (\ref{eq:Multi-valued-def}), no guards are satisfied. The conjunction of the guards for all levels is unsatisfiable, thus by definition of the part related to the guard in $f_{i,k}$~(\ref{eq:guard-booleanization}), we deduce that $f_{i,k}(w)=0$ by~(\ref{eq:final-formula-unitary-stepwise}), which is the expected result as $0$ is encoded by a Boolean profile filled with $0$ leading to $w'_{\sprt{y_i}_k}=0$ for all $k$. 
\end{enumerate} 

$f$ returns the appropriate result regarding a pair $w,w'$ encoding the pair $s,s'$. If $s \neq s'$ then there exists a Boolean support variable $\sprt{y_i}_k, 1 \leq k \leq |\sprt{y_i}|$ such that: $\psi(f_{i,k}(w) \cup w_{-\sprt{y_i}_k})= s'$, corresponding to the following condition: $f_{i,k}(w)\neq w_{\sprt{y_i}_k}$. Otherwise ($s = s'$) any index $k$ satisfies $\psi(f_{i,k}(w) \cup w_{-\sprt{y_i}_k})= s'$. Notice that this part is not sufficient for proving bisimulation, since we may have $f_{i,j}(w)= w'_{\sprt{y_i}_j} =w_{\sprt{y_i}_j}$ by definition of the asynchronous dynamics, thus also leading to a transition $w \xevo{\sprt{y_i}_j} w$ by definition. This transition does not simulate the transition $s \xevo{y_i} s'$ when $s \neq s$, motivating the proof of the bisimulation restricted to the reflexive reduction. However a Multi-valued self-loop ($s=s'$) is simulated by a self-loop in the Boolean network by construction of $f$. 

\medskip \noindent
\textit{B)  Property~(\ref{eq:main-fg}) is satisfied.}

 	Let $s \xevo{y_i} s'$ be a Multi-valued asynchronous transition such that $s \neq s'$,
 	then there exist  $w,w' \in \Bset_{\sprt{Y}}$ such that $\psi(w)=s,\psi(w')=s'$,  by construction of $f$ (A). Moreover, we have: 
 	$\psi(w) \neq \psi(w')$, leading to $w \neq w'$, as $\psi$ is a function. 
 Thus, a Boolean support variable $\sprt{y_i}_k$ verifies that $w_{\sprt{y_i}_k} \neq w'_{\sprt{y_i}_k}$, as $g$ and $f$ are neighbourhood preserving (\ref{eq:neighbourhood-preserving}).
 In this case,  we have: $w'=f_{i,k}(w) \cup w_{-\sprt{y_i}_k}$ from (A). Thus, we have the following equalities: 
 
 \begin{tabu} to \textwidth {X[r,$,0.5]  @{\;=\;} X[l,$,2] X[l,3]}
 	s' 	& \psi(w') &  by definition of $\psi$; \\	
 	    & \psi(f_{i,k}(w) \cup w_{-\sprt{y_i}_k}) & by construction of $f$ (A);\\
 	    & \psi(f_{i,k} \cup w_{\sprt{y_i}\setminus \sprt{y_i}_k \cup -\sprt{y_i}}) & from (\ref{eq:t1});\\
 		& \psi(f_{i,k} \cup w_{\sprt{y_i}\setminus \sprt{y_i}_k} \cup w_{-\sprt{y_i}}) & from (\ref{eq:support});\\
	& \psi(f_{i,k} \cup w_{\sprt{y_i}\setminus \sprt{y_i}_k})  \cup \psi(w_{-\sprt{y_i}}) & by (\ref{eq:support}), (\ref{eq:support-fun});\\
	& \psi(f_{i,k} \cup w_{\sprt{y_i}\setminus \sprt{y_i}_k})  \cup s_{-y_i} & as $s=\psi(w)$.
 \end{tabu}

\medskip
\noindent
As $s'= g_i(s) \cup s_{- y_i}$ by definition of a transition, we deduce by simplification of
 $s_{-y_i}$ that: $g_i(s) = g_i \circ \psi(w) = \psi(f_{i,k} \cup w_{\sprt{y_i} \setminus \sprt{y_i}_k})$,
concluding that Property~(\ref{eq:main-fg}) holds.

\medskip
\noindent
\textit{C) Bisimulation between reflexive reductions.}
It follows from (B), that we can set that Property~(\ref{eq:main-fg}) holds whenever $s_{y_i} \neq g_i(s)$. As  the 
asynchronous mode is local-to-support, and we always have $s \xevo{y_i} s' \implies s_{y_i}\neq 
s'_{y_i}= g_i(s)$ by reflexive reduction, 
we conclude by application of 
Theorem~\ref{thm:bisimulation-psi.f=g.psi-equivalence-local-to-support} that the  reflexive  
reduction of the Multi-valued and Boolean networks are bisimilar 
with respect to $\psi$.

\end{proof}

\begin{proof}[Proposition~\ref{prop:reach-admissible-region}]
We denote: $\mathbf{0}_m$ a Boolean state with $0$ for all variables in $m \subseteq \sprt{Y}$. 
	
	\medskip
	If a state is in the admissible region then it always reaches states in the admissible region, and only in the admissible region, by definition of bisimulation. 
	
	If the Boolean state $w \in \Bset_{\sprt{Y}}$ is outside of the admissible region, $w \notin \dom 
	\psi$, then it is not accounted for by the computation of admissibility of the guard condition, by
	definition of $\Psi_{\star \to 1}$. Therefore we have: $f_m(w)= \mathbf{0}_m, \forall m \in M$. 
	Thus, all the trajectories starting from $w$ successively cancel (set to 0) the states of the variables 
	of $m \in M$ whenever the result of the cancellation leads to a state outside the admissible 
	region; otherwise the proposition holds. As $ \bigcup_{m \in M} m = \sprt{Y}$ by definition of 
	a mode, the cancellation process terminates at state $\mathbf{0}_{\sprt{Y}}$, which is always in 
	the admissible region since $\mathbf{0}_{\sprt{y_i}}, \forall y_i \in Y,$ is the sole code for 
	the integer value 0, regardless of the variable $y_i$. 
	
	\medskip
	In conclusion, the trajectories starting from any $w \in \Bset_{\sprt{Y}}$ eventually end up in a state in the admissible region $\dom \psi$. 
\end{proof}

\pagebreak
 \bibliography{Multi-valued-bisimulation}
\bibliographystyle{plain}

\end{document}